\documentclass[12pt]{article}
\usepackage{amssymb}
\usepackage{mathtools}
\usepackage{hyperref}

\usepackage{fullpage}
\numberwithin{equation}{section}

\usepackage{titlesec}
\titlespacing*{\section}
{0pt}{0ex}{0ex}
\titlespacing*{\subsection}
{0pt}{0ex}{0ex}

\usepackage{graphicx,amsmath, bm}
\usepackage{natbib}

\usepackage{xcolor}         

\definecolor{DarkGreen}{rgb}{0.5,0.8,0.6}   
\definecolor{RGBblack}{rgb}{0.0,0.0,0.0}    

\definecolor{grau}{rgb}{0.8,0.8,0.8}

\newcommand{\chen}[1]{\color{orange}}

\def\stackover#1#2{\mathrel{\mathop{#2}\limits^{#1}}}
\newcommand{\iid}{\stackover{\mbox{\footnotesize i.i.d.}}{\sim}}

\makeatletter
\newcommand*{\rom}[1]{\expandafter\@slowromancap\romannumeral #1@}

\newcommand{\by}{\bm{y}}
\newcommand{\bY}{Y}

\newcommand{\bz}{\bm{z}}
\newcommand{\bw}{\bm{w}}

\newcommand{\bx}{\bm{x}}

\newcommand{\bs}{\bm{s}}

\newcommand{\bbh}{\bm{h}}
\renewcommand{\th}{\theta}
\newcommand{\sig}{\sigma}

\newcommand{\bmu}{\bm{\mu}}
\newcommand{\bbeta}{\bm{\beta}}

\newcommand{\Ga}{\mbox{Ga}}
\newcommand{\Dir}{\mbox{Dir}}

\newcommand{\DPP}{\mbox{DPP}}

\newcommand{\stau}{\sigma_q}

\newcommand{\up}{\mbox{\footnotesize up}}

\begin{document}

\title{Bayesian Inference for Latent Biologic Structure with Determinantal 
Point Processes (DPP)}

\author{Yanxun Xu \\
           {\small Department of Statistics and Data Sciences, The University of Texas at Austin, Austin, TX} \\
           {\small Department of Applied Mathematics and Statistics, Johns Hopkins University, Baltimore,  MD} \\
           Peter M\"uller \thanks{Address for correspondence: Department of Mathematics, The University of Texas at Austin, Austin, TX, 78712. Email: pmueller@math.utexas.edu}\\
          {\small   Department of Mathematics, The University of Texas at Austin, Austin, TX}\\
           Donatello Telesca \\
            {\small Department of Biostatistics, UCLA School of  Public Health, Los Angeles, CA} \\
}
\date{}
\maketitle

\newpage 
\begin{abstract}
We discuss the use of the determinantal point process (DPP) as a prior
for latent structure in biomedical applications, where inference often
centers on the interpretation of latent features as biologically or
clinically meaningful structure.
Typical examples include mixture models, when the terms of the mixture
are meant to represent clinically meaningful subpopulations (of
patients, genes, etc.).
Another class of examples are feature allocation models. We
propose the DPP prior as a repulsive prior on latent mixture
components  in the first example, and as prior on
feature-specific parameters in the second case.
 We argue that the DPP is in general an attractive prior model 
for latent structure when biologically relevant
interpretation of such structure is desired.
We illustrate the advantages of DPP prior in three case studies,
including inference in mixture models for magnetic resonance images
(MRI) and for protein expression, and a feature allocation model for
gene expression using data from The Cancer Genome Atlas.  An important
part of our argument are efficient and straightforward posterior
simulation methods. We implement a variation of reversible jump Markov
chain Monte Carlo simulation for inference under the DPP prior, using
a density with respect to the unit rate Poisson process.

\end{abstract}

\clearpage
\newpage
\section{Introduction}
\label{sec:intro}
Independent priors for latent structure are almost never
 appropriate in biomedical inference. Nevertheless, they are widely
 used, simply for technical convenience and the lack of
 good alternatives. In this paper we argue for an attractive class of such alternative
 models in typical inference problems in biostatistics and bioinformatics.  
  
We discuss the use of the determinantal point process (DPP) for
modeling latent biologic structure. In particular, we focus on mixture
models and feature allocation problems, when the latent components 
are to be interpreted as biologically meaningful structure. For example, in the
case of a mixture model, we might want to interpret components of a
mixture as clinically meaningful patient subpopulations. Similarly, when using
feature allocation to model latent tumor cell subpopulations we might want
to interpret the features as substantially distinct subclones  \citep{xu2015mad}.
In both cases, an important aspect of the problem is the preference
for the latent 
elements being diverse. Such inference is poorly
formalized by traditionally used  independent priors.
We suggest the DPP prior as an attractive alternative
to implement repulsive priors.
The use of the DPP for mixture models is not novel. It was originally
proposed in \cite{affandi2013approximate}, but remains curiously
under-used in biomedical literature.
The contribution of the following discussion is the 
emphasis on problems with small to moderate size mixtures, the
extension to inference for general latent structure, and the detailed
posterior Markov chain Monte Carlo (MCMC) scheme, including easy to implement transdimensional posterior
simulation across different size latent structures.

For the moment we restrict attention to parametric
mixture models, to be
specific and also because such models are perhaps the most common
models for latent structure in biomedical applications.
For example, popular Bayesian models for clustering and inference on
patient subpopulations are variations of the following model. Let $y_i$
denote a response for the $i$-th patient. We assume
\begin{equation}
  y_i \sim \sum_{h=1}^H w_h\, p(y_i \mid \mu_h),
\label{mix}
\end{equation}
 $i=1,\ldots,n$,  
including possibly $H=\infty$. 
The component-specific sampling model $p(y_i \mid \mu_h)$ could be, for
example, a survival model with parameters $\mu_h$, possibly including a
regression on patient covariates.
The use of independent priors for component-specific parameters $\mu_h$
then gives rise to concerns about over-fitting
that generates redundant mixture components with similar parameters,
leading to unnecessarily complex models and poor interpretability.
In particular, such over-fit  compromises the interpretation of the mixture
components as biologically meaningful structure.
\cite{rousseau2011asymptotic} argued that such concerns were 
asymptotically partially mitigated with carefully chosen priors.
Alternatively, \cite{petralia2012repulsive} proposed a class of
repulsive priors for mixture components. The proposed repulsive prior
was based on a distance metric in which small distances were
penalized. They showed that using repulsive priors on location
parameters resulted in better separated clusters, while keeping the
density estimation accurate. However, posterior computations are
complex and do not readily extend to high dimensional cases.

An alternative interpretation of \eqref{mix} is as a mixture,
$y_i \sim \int p(y_i \mid \mu)\, dG(\mu)$, with respect to a
discrete probability measure $G= \sum w_h\delta_{\mu_h}$. If the model is
completed with a Dirichlet process (DP) prior on $G$ the popular DP
mixture model is obtained. See, for example, \cite{Ghoshal:10} for a review
of such nonparametric Bayesian models. 
Importantly, the DP prior includes independence across $\mu_h$. 

For later reference note that \eqref{mix} can be equivalently written
as a hierarchical model with latent indicators $s_i$,
\begin{equation}
  y_i \mid s_i=k  \sim  p(y_i \mid \mu_k)
\mbox{ and }
  p(s_i=k)  =  w_k.
  \label{eq:likl}
\end{equation}
Interpreting the latent indicators as cluster membership indicators,
model \eqref{eq:likl} includes inference on a random partition
$\bs=(s_1,\ldots,s_n)$ of $\{1,\ldots,n\}$.
Let $S_k = \{i:\, s_i=k\}$ denote the $k$-th cluster.
To avoid the notion of empty clusters, that is $|S_k|=0$, we re-arrange
the indexing of the $\mu_h$ to start with
$h=1,\ldots,K$ corresponding to non-empty clusters. 
Again, an independent prior on the cluster-specific parameters
$\mu_h$ complicates a meaningful interpretation of posterior inference
on the random partition $\bs$.

In this paper we argue for an alternative model that replaces the
independent prior on $\mu_h$ by the repulsive DPP
\citep{macchi1975coincidence}.
Recent reviews of the DPP appear in
\cite{lavancier:15} 
and, specifically for finite state
spaces, in \cite{kulesza2012determinantal}.
The use of the DPP as a prior for statistical inference in mixture
models, we believe, is first discussed in
\cite{affandi2013approximate}. 
The main contributions of this paper are
the recognition of the DPP as an attractive prior for latent features
in general latent structure models, including mixture models and
latent feature allocation as specific examples; 
the discussion of DPP mixtures specifically when one wants to
interpret latent structure as biologically meaningful
features;
and an easily implemented posterior simulation scheme for
a moderate number of latent structures, as is typical for biomedical
inference problems.
Posterior simulation is implemented as a variation of reversible
jump (RJ) MCMC simulation 
\citep{green:95} for a density with respect to the unit rate Poisson
process.

\section{Motivating Example}
\begin{figure}[!ht]
\centering
\includegraphics[scale=0.7]{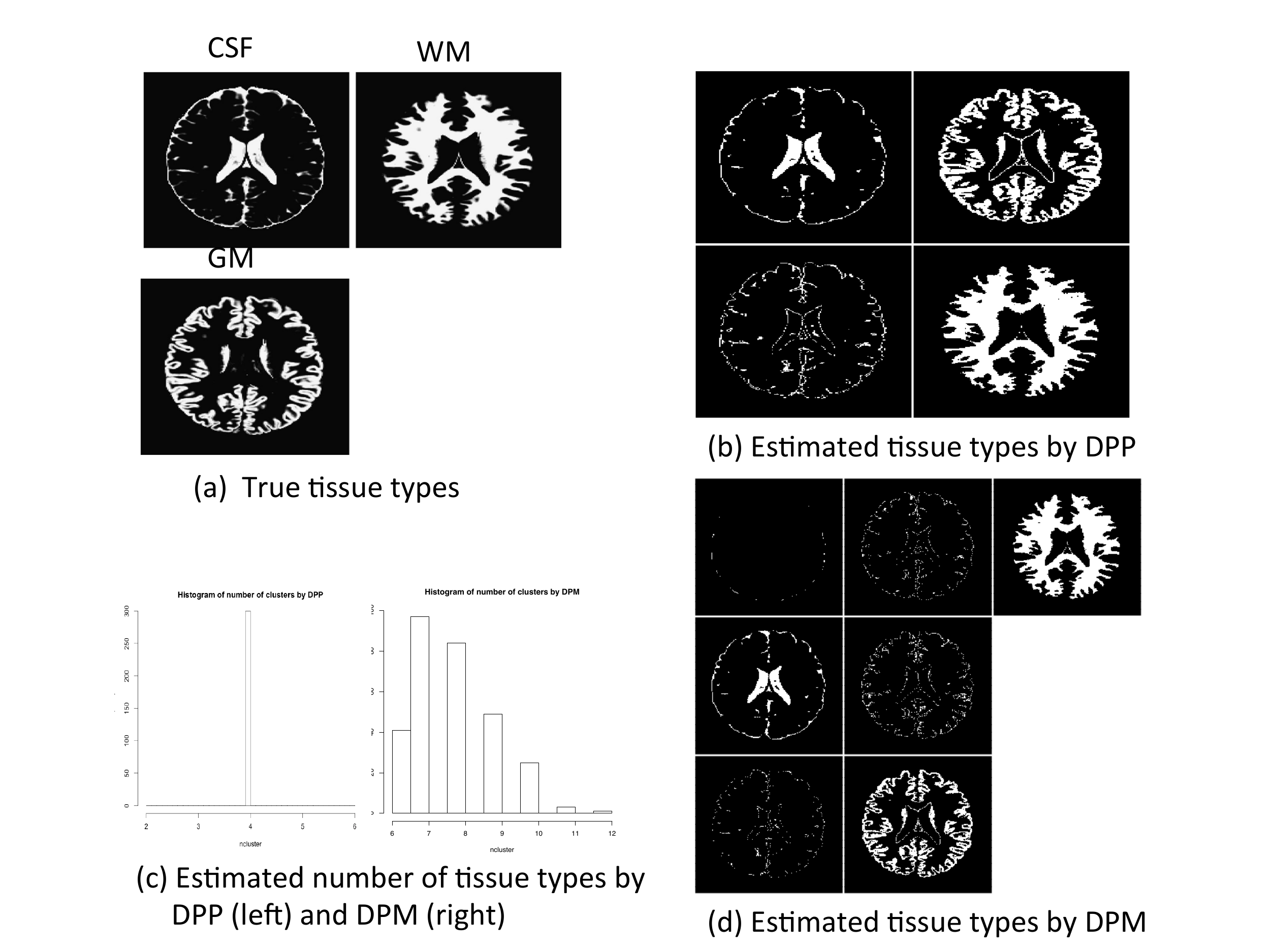}
\caption{BrainWeb images. Panel (a) shows the three true tissue
  types: CSF,  WM and GM. Panel (b) shows the estimated tissue types
  under the DPP prior. Panel (c) shows the estimated number of tissue
  types by DPP (left) and DPM   (right).  Panel (d) shows the
  estimated tissue types under the DPM prior. }  
\label{fig:image}
\end{figure}

Magnetic resonance imaging (MRI) is an effective technique for
studying the human brain. 
For example, MRI volume
estimates of white matter (WM), gray matter (GM), cerebrospinal fluid
(CSF), and their spatial distribution help the diagnosis of
degenerative brain illnesses, like Alzheimer's disease
\citep{decarli1992method}. Therefore, accurate clustering of MRI data
according to tissue types is vital to diagnosis and clinical research.
To illustrate, we download a sample of simulated
imaging data from BrainWeb \citep{cocosco1997brainweb} for slice
number 92. Figure \ref{fig:image}a depicts the ground truth
components for CSF, WM and GM. We implement inference under model-based
clustering with a DPP prior and a similar model
 based on the widely used Dirichlet process mixture (DPM)
model. Model details will be discussed later. For the moment we only 
intend to highlight the nature of the inference under the two models to
motivate the upcoming discussion.
Figure \ref{fig:image}c shows the posterior distribution $p(K \mid
data)$ on the number of clusters estimated under the
DPP prior (left panel) and the DPM prior (right panel).  As
shown in Figure \ref{fig:image}b, the DPP clustering model identifies
four clusters, 
three of which match the simulation truth and the last
one is simulated noise. In contrast, inference under the
DPM model finds seven clusters,
only three of them having a meaningful explanation.

\section{Determinantal Point Process (DPP)}
\subsection{Definition}
The DPP defines a point process on $S \subseteq \Re^D$,
that is, a random point configuration $X=\{x_1,\ldots,x_K\}$ with $x_k
\in S$. 
We first define it for a finite state space, $S = \{\omega_1,\ldots,\omega_N\}$.
Let $C$ denote an $(N \times N)$ positive semidefinite matrix,
constructed, for example, as $C_{ij}=C(\omega_i,\omega_j)$
with a covariance function $C(\omega_i, \omega_j)$.
Let $C_A$ denote the submatrix of rows and columns indicated by $A
\subseteq S$.
In later applications we will identify $x_k$ as 
$\mu_k$ in mixture models like \eqref{eq:likl}, latent feature allocations etc. 
For the moment we consider a generic random point
configuration $X$, defined as 
\begin{equation}
  p(X=A) = \det(C_A)/\det(C+I)
\label{DPP-finite}
\end{equation}
as a probability distribution on the $2^N$ possible point
configurations $X \subset S$. 
This defines a subclass of DPPs known as L-ensembles. 
It is easy to see why \eqref{DPP-finite} defines a repulsive point process if
one interprets the determinant as the volume of a parallelotope 
spanned by the column vectors of $C_A$. Equal or similar column
vectors span less volume than very diverse ones.
Equation \eqref{DPP-finite} can be shown to imply the marginal probabilities
\begin{equation}
  p(A \subseteq X) \propto \det(M_A)
\label{DPP-marginal-finite}
\end{equation}
for $M = C(I+C)^{-1}$
 \citep{kulesza2012determinantal}, where $M_A$ is a submatrix of $M$. 
Equation \eqref{DPP-marginal-finite} defines a DPP on a finite state
space $S$. Every L-ensemble is a DPP. But not every DPP is an
L-ensemble. For singular $(I-M)$ we can define
\eqref{DPP-marginal-finite}, but not \eqref{DPP-finite}.
A good review of DPP models for finite state spaces,
including the derivation of the normalizing constant in
\eqref{DPP-finite} appears in
\cite{kulesza2012determinantal}.

For a continuous  state space $S \subseteq \Re^D$, we define an 
L-ensemble by a density $f(X)$  
with respect to the unit rate Poisson process as
\begin{equation}
  f(X) = \det(C_X)/\prod_{h=1}^\infty (\lambda_h+1).
\label{DPP}
\end{equation}
for $X=\{x_1,\ldots,x_K\}$.
As before, $C_X$ is a $(K \times K)$ matrix with $(i,j)$ entry 
defined by a continuous covariance function $C(x_i,x_j)$.
The $\lambda_h$'s are the eigenvalues of the associated kernel operator 
$\int_SC(x, y)h(y)dy$. 
Similar to the case of a finite state space,
it is possible to generalize \eqref{DPP} to the slightly
larger class of DPP models \citep{lavancier:15}.
However, for the rest of this discussion we will consider L-ensembles and
work with the kernel $C(x_i,x_j)$ only.

For continuous DPP kernels, the eigenvalues $\lambda_h$ are generally
unknown except for a few kernels such as a squared exponential kernel.  
Several numerical methods are used to approximate eigenvalues and
corresponding eigenfunctions
\citep{lavancier:15}.
We build on \cite{kulesza2010structured} and decompose the kernel function $C$ as
\begin{equation}
  C(x, x') = q(x)c(x,x')q(x')   \ \ \ \ \ \ \ \ 
  x, x' \in {\cal X} \ \ {\mbox
  {and}} \ \ c(x, x)=1, 
\label{kernel}
\end{equation}
where $q(x)$ is  the quality function and $c(x,y)$ is the similarity
kernel. 
For a multivariate $\bx=(x_1,\ldots,x_D) \in \Re^D$ 
we use
$$
 q(\bx)= \prod_{d=1}^D
 \frac{1}{\sqrt{2\pi}\stau}\exp\left\{-\frac{x_{d}^2}{2\stau^2} \right\}
\mbox{ and }
   c(\bx, \bx') = 
    \exp\left\{-\sum_{d=1}^D\frac{(x_{d}-x'_{d})^2}{\theta^2}\right\},
$$
for which \cite{zhu1997gaussian} gives
analytic results for the eigenvalues and eigenfunctions.
%
Eigenvalues $\lambda_{\bbh}$ are given by:
\begin{equation}
   \lambda_{\bbh} = \prod_{d=1}^D
   \sqrt{\frac{2a}{a+b+c}}\left(\frac{b}{a+b+c}\right)^{h_d-1},
\label{eq:lam}
\end{equation}
where $\bbh=(h_1, \dots, h_D)$ is a multivariate index,
$a=\frac{1}{4\stau^2}$, 
$b = \frac{1}{\th^2}$, and 
$c= \sqrt{a^2+2ab}$. 
Here, 
$\th$ and $\stau$ are hyperparameters
that define the kernel function.

We write $X \sim \DPP(C, \th, \stau)$ for 
$X=\{x_1,\ldots,x_K\}$ generated by a DPP model with a kernel
function $C(\cdot,\cdot)$ that is indexed with parameters $\th,\stau$,
and we write $\DPP(C)$ when $C(\cdot,\cdot)$ involves no unknown
hyperparameters.  

\subsection{Posterior Simulation}
\label{sec:mcmc}
Later we will use the DPP as prior probability model for latent
structure, including latent clustering and feature allocation.
In both cases, an important step in the posterior simulation will be
a transition probability to change the number of atoms in the DPP.
We discuss a reversible jump (RJ) scheme to implement such 
transition probabilities
using the density \eqref{DPP} with respect to the unit rate Poisson
process. 
Let $\Omega_K$ denote the $\sigma$-algebra for size $K$ point
configurations, and $\Omega = \bigcup_{K=0}^\infty \Omega_K$.
We define an MCMC transition probability that allows a move from 
$F_K \in \Omega_K$ to $F_{K+1} \in \Omega_{K+1}$ or $F_{K-1} \in \Omega_{K-1}$.
The algorithm combines the MCMC simulation for a
point process from \cite{Geyer&Moller94} with the deterministic
transformation that is included in the reversible jump (RJ) scheme of
\cite{green:95}.
The construction parallels the construction of \cite{green:95}, with
only a minor variation that is needed to reduce the integral with
respect to the unit rate Poisson process to an integral with respect
to Lebesgue.

Assume the current state is $x=\{x_1,\ldots,x_K\}$ and
we consider two transition probabilities, $P_u(dy \mid x)$
which proposes a move to a size $K+1$ point configuration (``up'' move)
and $P_d(dx \mid y)$ which proposes a move to a size $K-1$ point
configuration (``down'' move). For example, $P_u$ could be proposing
to split  one of the atoms in $x$ into two daughters, thereby
incrementing $K$ by one; and $P_d$ could involve merging
two points in $x$. 
Let $q(x)$ denote the probability of choosing $P_u$, and let
 $A_u(x,y)$ and $A_d(y, x)$ 
denote the acceptance probability for a proposal $y$. 
Finally, let $f(x)$ denote the density
\eqref{DPP} with respect to the unit rate
Poisson process $\mu(\cdot)$.
The detailed balance condition becomes
\begin{multline}
   \int_{F_{K+1}} (1-q(y))
    \left[ \int_{F_K} A_d(y,x) P_d(dx \mid y) \right]\;
  f(y) d\mu(y) 
= \\
   \int_{F_K} q(x)
    \left[ \int_{F_{K+1}} A_u(x,y) P_u(dy \mid x) \right]\;
  f(x) d\mu(x).
\label{DB}
\end{multline}
Assume that there are $n_{\up}(x)$ possible up moves, $j=1,\ldots,n_{\up}(x)$. 
For example, if the up move involves splitting one of the atoms 
of the size $K$ point configuration $x$,
we could choose one of the $n_{\up}(x)=K$ points to split.
Let $q_{uj}(x)$ denote the probability of selecting the $j$-th transition
probability. That is
$P_{u}(dy \mid x) = \sum_{j} q_{uj}(x) P_{uj}(dy \mid x)$.
Similarly, 
$P_{d}(dx \mid y) = \sum_{j} q_{dj}(y) P_{dj}(dx \mid y)$.
A sufficient condition for detailed balance is that equation \eqref{DB}
holds for pairs of moves, $P_{uj}, P_{dj'}$ 
that are defined and linked in the following sense.
We assume that $P_{uj}$ is constructively defined by (i) 
generating an auxiliary variable $u \sim q_u(u \mid x)$; (ii) a
deterministic, invertible transformation $y = T(x,u)$;
and (iii) the matching down move $P_{dj'}$ is defined by
$x = T_1^{-1}(y)$. Here $T^{-1}_1(y)$ denotes the first element of
$T^{-1}(y)=(x,u)$. 
The detailed balance condition becomes
\begin{multline}
   \int (1-q(y))q_{dj'}(y)
    \left[ A_d(y,x) 
       I\{x=T^{-1}_1(y) \in F_K;\; y \in F_{K+1}\} \right]\;
  f(y) d\mu(y) 
= \\
  \int  q(x) q_{uj}(x)
    \left[ \int A_u(x,y) 
       I\{x \in F_K;\; y = T(x,u) \in F_{K+1}\} q_u(u \mid x) du
    \right]\;
  f(x) d\mu(x).
\nonumber 
\end{multline}
We replaced the range of integration by an indicator for $x \in F_K$
and $y \in F_{K+1}$.
Next we use 
$\int_{F_K} h(x) d\mu(x) = 
\frac{e^{-|S|}}{K!}\, \int h(x) I(x \in F_K)\, dx_1\cdots dx_K$. 
That is, a unit rate Poisson process restricted to size $K$ point
configurations looks exactly like $K$ i.i.d. uniform random variables on
$S$ \citep{kingman:92}.
The extra factor
 $e^{-|S|} |S|^K/K!$ 
 arises from the probability of a size $K$
point configuration.
Note that $x=\{x_1,\ldots,x_K\}$ remains the (unordered) point configuration.
We get
\begin{multline}
   \int (1-q(y))q_{dj'}(y)
    \left[ A_d(y,x) 
       I\{x \in F_k;\; y \in F_{K+1}\} \right]\;
   \frac{f(y)}{(K+1)!}\, dy_1\cdots dy_{K+1} = \\
   \int q(x)q_{uj}(x)
    \left[ \int A_u(x,y) 
       I\{x \in F_k;\; y \in F_{K+1}\} q_u(u \mid x) du
    \right]\;
  \frac{f(x)}{K!}\, dx_1 \cdots dx_K,
\label{DB3}
\end{multline}
still using $x=T^{-1}_1(y)$ on the left and 
$y=T(x,u)$ on the right hand side.
Finally, we use a change of variables,  
substituting
$dy_1 \cdots dy_{K+1}$ by $dx_1\cdots dx_K du |J|$ with the Jacobian 
$J = \partial T/\partial x_1\cdots\partial x_K \partial u$.
A sufficient condition for \eqref{DB3} 
is the equality of the two integrands,
$
  (1-q(y)) q_{dj'}(y)\,
    A_d(y,x)\;
   \frac{f(y)}{K+1} |J| = $
$  q(x) q_{uj}(x)\,
       A_u(x,y)\;
       q_u(u \mid x)
  f(x), 
$
for  $x \in F_K$ and $y = T(x,u) \in F_{K+1}$.
The condition is verified for
$A_u(x,y) = \min\{1, \rho(x,y)\}$ and $A_d(y,x)=\min\{1,
1/\rho(x,y)\}$
with
\begin{equation}
  \rho(x,y) = \frac{f(y)}{(K+1)f(x)}\,
              \frac{1-q(y)}{q(x)}
              \frac{q_{dj'}(y)}{q_{uj}(x)}
              \frac{1}{q_u(u \mid x)}
              |J|.
\label{rho}
\end{equation}
Acceptance probability \eqref{rho} defines essentially the RJ 
 algorithm of \cite{green:95}.
The only minor difference is the extra step of representing the
probability of a point configuration with respect to the unit rate
Poisson process by a probability of the ordered $K$-tuple
$(x_1,\ldots,x_K)$.
\cite{Geyer&Moller94} use the latter for a birth and death Markov
chain Monte Carlo, and without the deterministic transformation.
For posterior simulation conditional on data $y \sim p(y \mid x,\th)$
multiply with an additional likelihood ratio in \eqref{rho}. Here $\th$ are
additional parameters in the sampling model, beyond $x$.

In summary, we have shown that the density with respect to the 
unit rate Poisson process can be used to construct a RJ MCMC,
essentially as if it were a density with respect to Lebesgue.
A similar argument holds for Metropolis-Hastings transition
probabilities, without a change in the size of the point configuration $x$.

\section{DPP Clustering}
\label{sec:cluster}
\subsection{Motivation and Model}
Clustering is fundamental to exploratory analysis of bioinformatics
data. For instance, elucidating patterns of gene expression 
and identifying 
sets of genes that behave similarly under certain
biologic conditions is important in the study of functional genomics and proteomics. Clustering also can be applied to
develop targeted therapies. We first cluster the patient samples into
several subgroups based on protein activation (or some other 
patient baseline characteristics),
then correlate patient clusters with overall survival and
investigate subgroup-specific therapies.
These and similar applications in biomedical inference
motivate the following model.  

We start with a mixture of normals sampling model, 
as it is widely used in clustering and density estimation.
Here we show simulation with the univariate sampling model 
\eqref{eq:likl}. In Web Appendices A and B we show a straightforward extension to a
multivariate mixture, including a brief simulation study. 
We assume that data 
$\by_n=\{y_i\}_{i=1}^n$ are generated from 
$
  y_i \sim \sum_{k=1}^K w_k N(\cdot\mid \mu_k, \sigma_k^2),
$ 
$i=1, \dots, n$, with unknown $K$. 
This is a special case of \eqref{mix} 
with random $K$ and with a normal kernel $p(y_i \mid
\mu_k,\sig^2_k)$.
The model implies a prior for a random
partition $\bs=(s_1,\ldots,s_n)$, as in \eqref{eq:likl}.
Often the inference goal is to identify latent clusters 
$S_k=\{i:\; s_i=k\}$ that
correspond to meaningful biologic conditions or
to identify subpopulations that are sufficiently diverse to  be considered for
different clinical decisions such as treatment allocation.
The protein data analysis for kidney cancer patients, 
in Section \ref{sec:KIRC}, is a typical example.
In such problems an independent prior on $\mu_k$ has the undesirable
feature of allowing for very similar, or even identical (in the case
of a discrete parameter space) $\mu_k$. 
To interpret different terms in the mixture as meaningful
structure in the population, we prefer a repulsive prior on 
the $\mu_k$, that is, a probability model that favors a priori very
distinct values $\mu_k$. 
The repulsive property and the
relative computational simplicity make the
DPP an appealing choice. \cite{kwok2012priors} applied  the DPP as
repulsive prior in latent variable models. However, for lack
of efficient computational algorithms 
their method was restricted to MAP (maximum a posterior) inference.
\cite{affandi2013approximate}
proposed a Gibbs sampling technique for inference with DPP priors 
under fixed $K$ ($K$-DPP).
The posterior simulation scheme from Web Appendix A
allows us to implement inference under an unconstrained DPP prior,
including a random size ($K$) point configuration.

\paragraph*{The DPP mixture model.}
We complete the sampling model \eqref{eq:likl} with
a DPP prior on the cluster-specific parameters $\mu_k$:
\begin{gather}
  y_i \mid s_i=k  \sim  p(y_i \mid \mu_k)
  \mbox{ and }
  p(s_i=k)  =  w_k. \nonumber \\
  \bmu = \{\mu_1, \dots, \mu_K\} \sim \DPP(C, \theta, \stau),
\label{eq:prior}
\end{gather}
using the kernel function in \eqref{kernel}. Recall that $\th,\stau$
are hyperparameters in the definition of $C$.
Finally  we use hyperpriors 
$
  \bw \mid K, \delta \sim \Dir(\delta, \dots, \delta),~~  
  1/\sigma_k^2 \sim \Ga(a_0, b_0),~~
  \theta \sim N(a_1, b_1^2), \mbox{ and }
  \stau \sim N(a_2, b_2^2). 
$
Here $\Ga(a,b)$ refers to a Gamma distribution with mean $a/b$. 
The model can be easily extended to multivariate responses using
multivariate normal and inverse-Wishart priors.
%
Posterior inference is carried out using MCMC simulations. Details are shown in Web Appendix A.

\begin{figure}[!h]
\centering
\begin{tabular}{cc}
\includegraphics[scale=0.3]{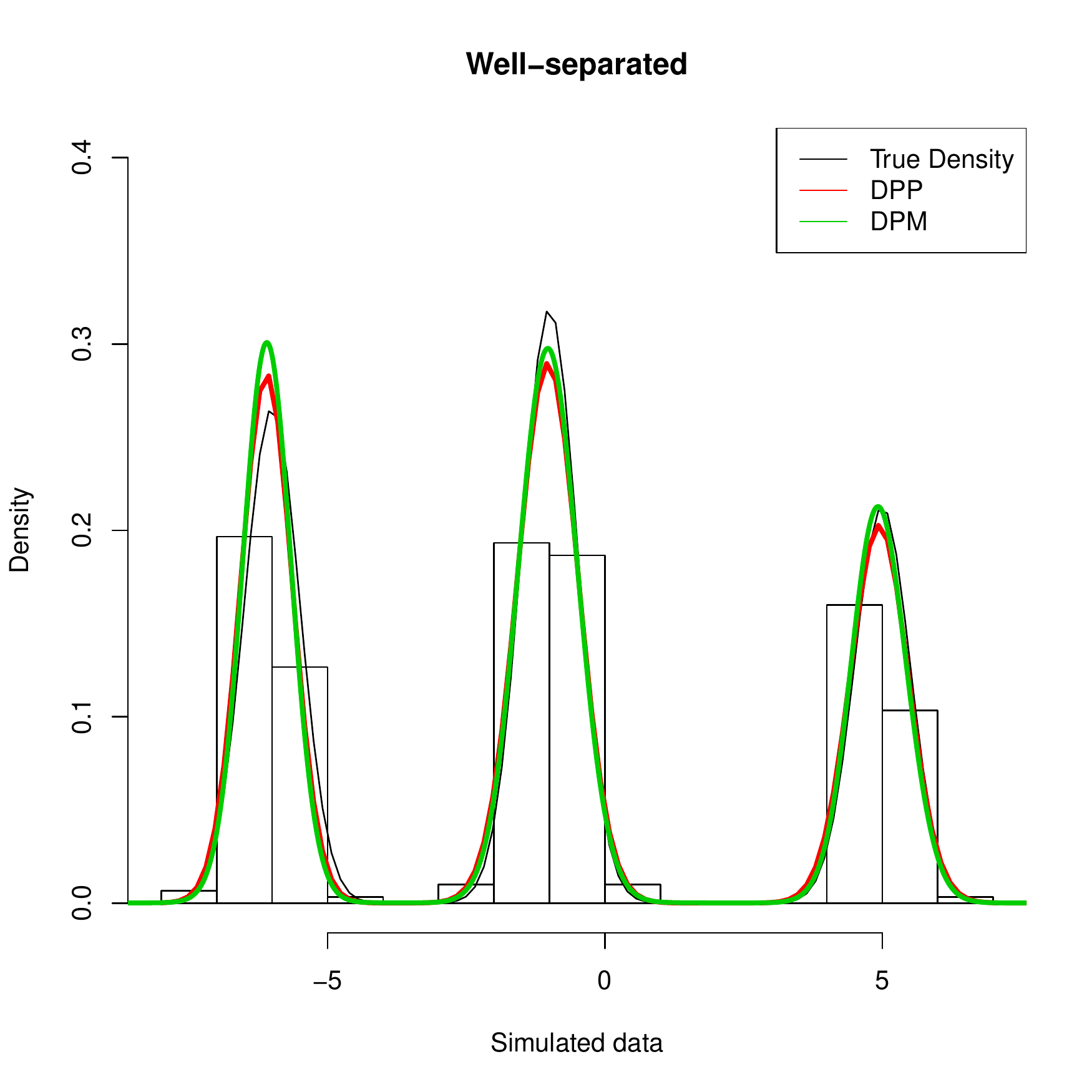}&\includegraphics[scale=0.3]{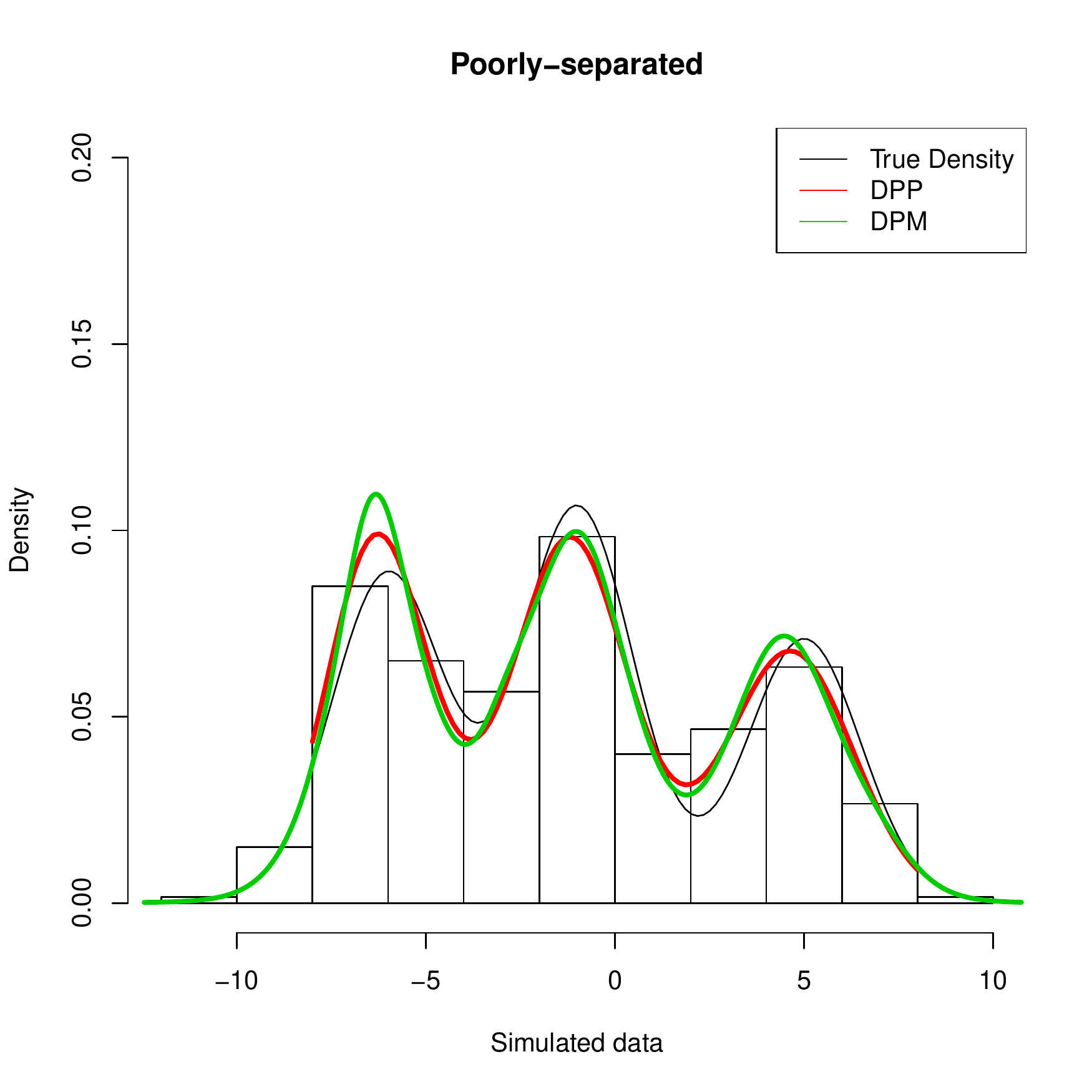}\\
\includegraphics[scale=0.3]{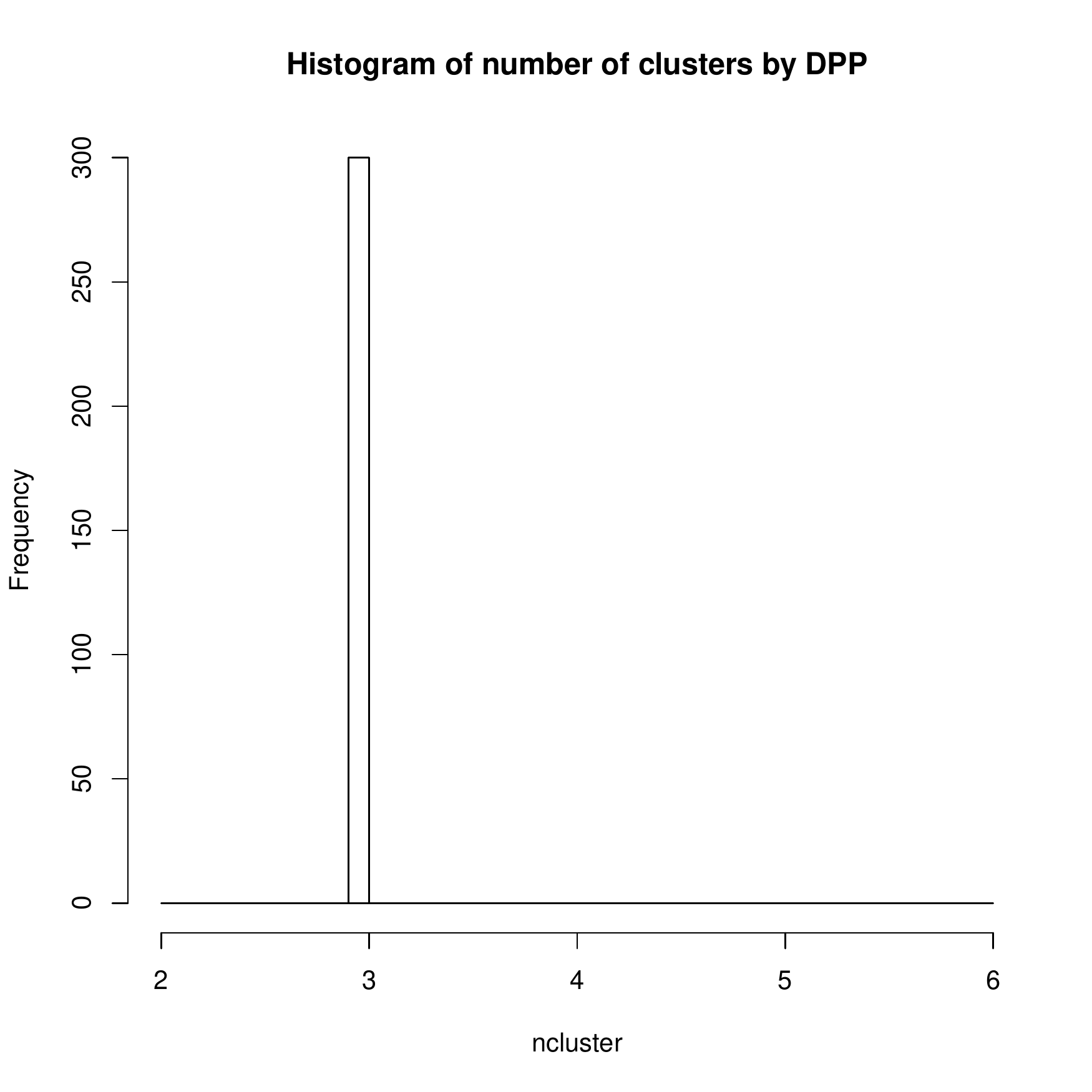}&\includegraphics[scale=0.3]{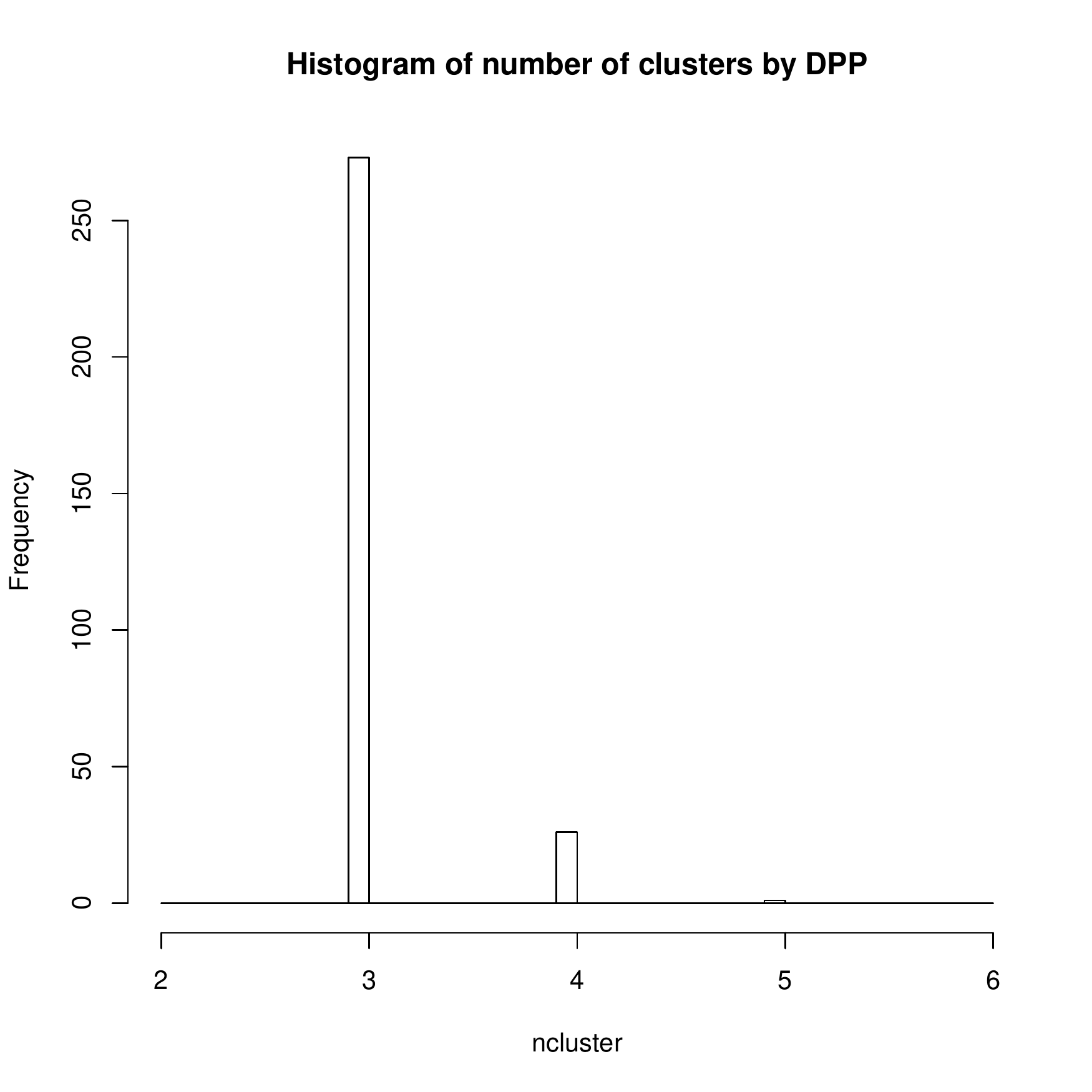}\\
\includegraphics[scale=0.3]{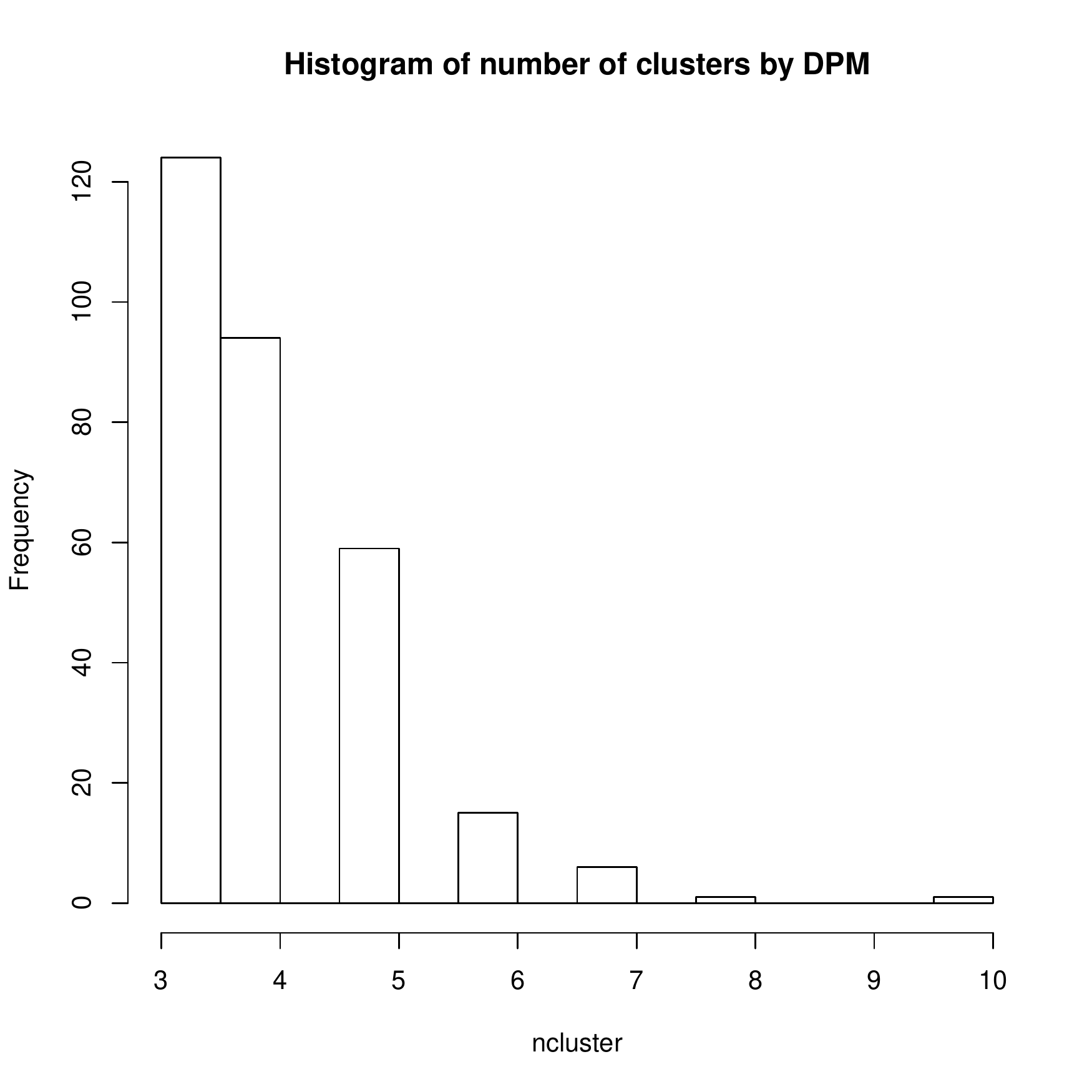}&\includegraphics[scale=0.3]{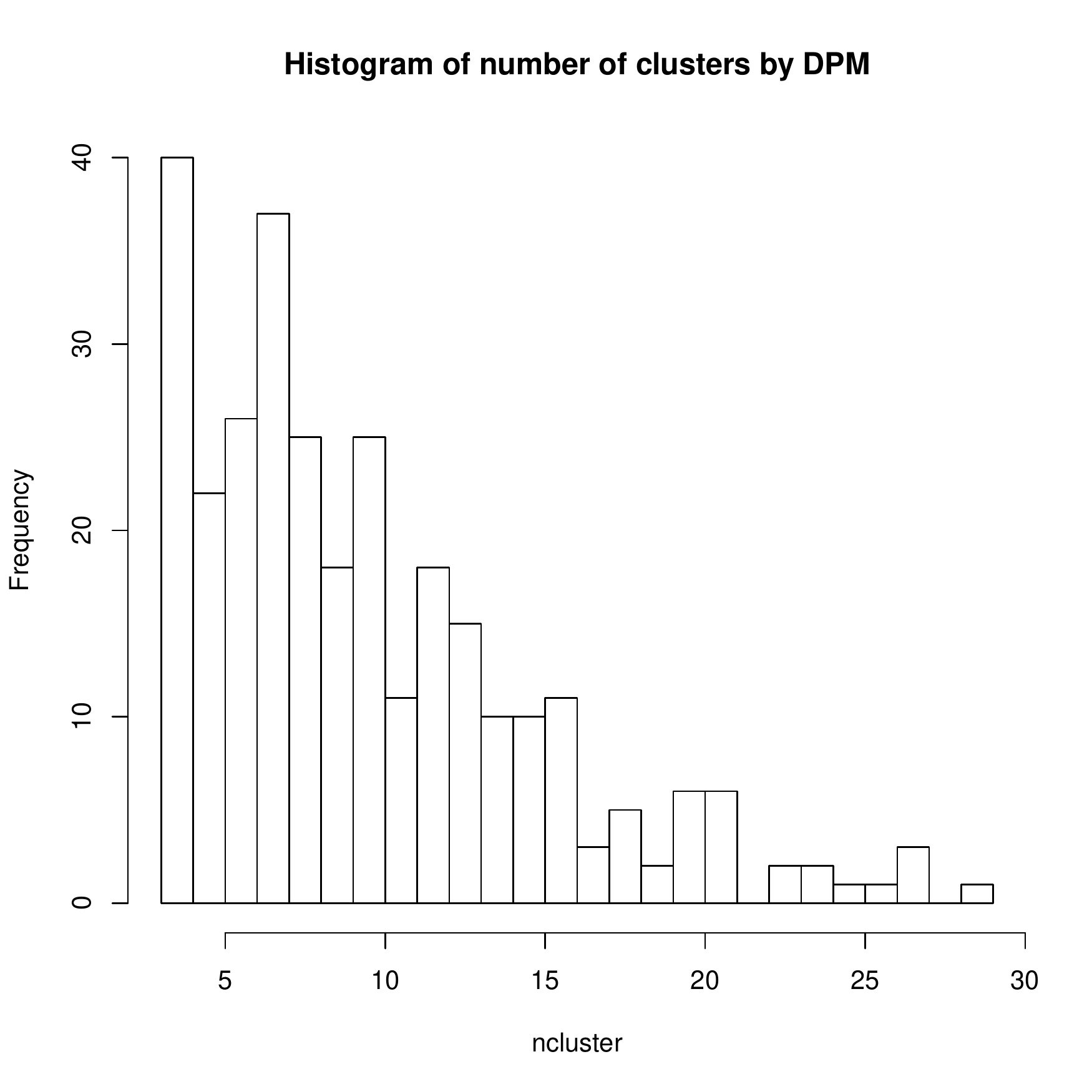}\\
$\sigma = 0.5$&$\sigma = 1.5$\\
\end{tabular}
\caption{Simulation: DPP mixture model.
  The upper panel shows the histograms of two simulated
  datasets with true density (black), estimated density by DPP prior
  (red) and DPM prior (green). The lower panels present the histograms of
  the estimated number of  clusters by DPP prior (2nd row) and DPM prior 
  (3rd row). } 
\label{fig:simu1}
\end{figure}

\paragraph*{Two simulation studies.}
We carry out two simulation studies 
to evaluate the performance of the repulsive DPP prior in
clustering and density estimation, 
with both univariate and multivariate responses. 
Results are summarized in Figure \ref{fig:simu1}.
Details of the simulation setup and more results are shown in
Web Appendix B. See there also for more discussion of Figure
\ref{fig:simu1} 
and for a statement of the multivariate version of the
DPP mixture model. 
The results show that
the DPP prior leads to a sparser representation with interpretable
clusters compared to DPM, while maintaining a good fit for the
density estimate, making it a preferable prior model for applications
where such parsimony is desired.

\subsection{KIRC Protein Data Analysis}
\label{sec:KIRC}

\begin{figure}[!h]
\centering
\includegraphics[scale=.65]{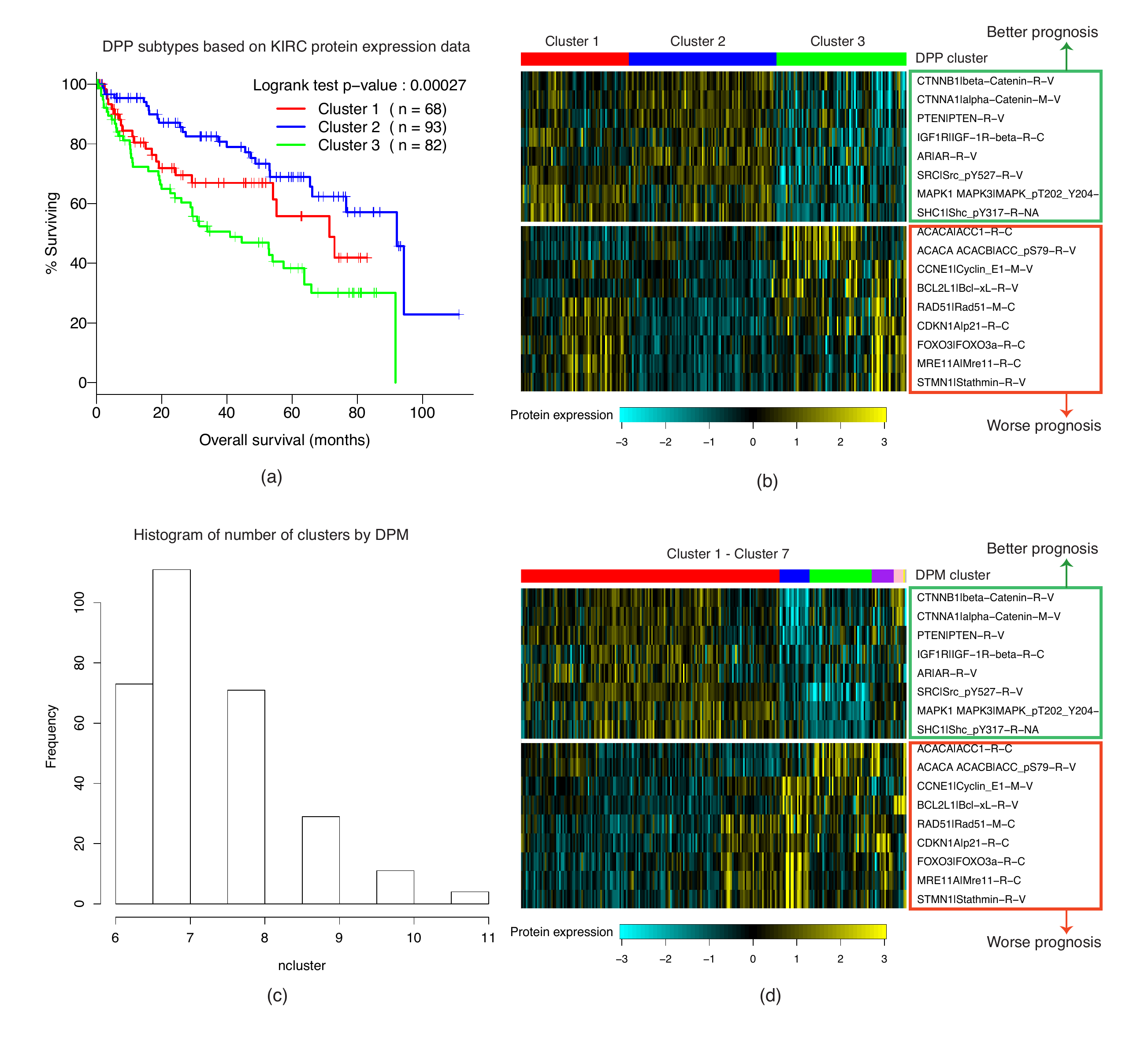}
\caption{KIRC data.
  Panel (a) shows a Kaplan-Meier plot of overall survival in the KIRC core
  set stratified by three clusters identified under DPP prior. 
  Panel (b) shows the top
  differentially expressed protein markers among three DPP clusters. Columns correspond to patients, rows correspond
    to proteins.
  Panel (c) is the histogram of the number of clusters identified under 
  DPM prior. 
  Panel (d) shows a heatmap of seven DPM clusters for top differentially expressed
  protein markers. The sizes of the seven clusters are 163, 19, 39, 14,
  6, 1 and 1, respectively. }
\label{fig:real2}
\end{figure}

%
We implement inference under the proposed DPP mixture model
for protein expression data
 from \cite{yuan2014assessing}
 with $n=243$ samples and $D=17$ protein
markers for kidney renal clear cell carcinoma (KIRC).
See equation (4) in Web Appendix A.2 for a statement of the DPP 
   mixture model (4.1) with a multivariate normal kernel $p(y_i \mid  \mu_k)$. 
Inference goals include 
correlating protein expression with patients' overall survival.  
The $n=243$ KIRC samples are classified into three clusters by the
proposed DPP mixture model. As shown in Figure \ref{fig:real2}a,
patients stratified by these three DPP groups exhibit very distinct
survival patterns ($p$-value under a log-rank test is $p=0.00027$).
Proteins that are correlated with better prognosis are relatively elevated in cluster 2
(the best survival group) while the proteins correlated with worst
survival are relatively elevated in clusters 1 and 3 (especially
cluster 3, the worst survival group)
(Figure \ref{fig:real2}b).
These results suggest that
inference under the DPP prior can successfully classify patients into
biologically meaningful groups based on molecular profiles.

In contrast, under the DPM mixture model, $K=7$ clusters are identified,
estimated as the mode of $p(K \mid \by)$ shown in Figure
\ref{fig:real2}c. Most of the 
seven clusters have small size ($<$ 20) while 2/3 of the samples
are allocated to one cluster (the red bar in Figure
\ref{fig:real2}d). 
The clusters are not easily interpreted 
(Figure  \ref{fig:real2}d).

In summary, inference under the DPP prior provides fewer clusters and
gives more interpretable results in molecular profile-based classifications
than inference under a comparable DPM prior.

\section{A DPP Feature Allocation Model}
\label{sec:latent}
\subsection{Motivation and Model}
Breast cancer is a heterogeneous disease in terms of molecular
alterations and clinical responses. Gene expression profiling
can provide valuable information for understanding this complexity
and consequently for predicting clinical outcomes.
Here we consider a study reported in
\cite{chen2013phylogenetic} who  aim to  characterize gene expression profiles
by a small number of underlying distinct molecular drivers.
These latent molecular drivers  should be  linked to different
subsets of samples.
This motivates us to propose the model below which formalizes 
this preference by using a DPP prior for  the pattern of how molecular
drivers  (the columns of the matrix $Z$ below) are linked to
samples (rows of $Z$). 
%
Let $\bY$ denote the observed $n\times S$ data matrix with rows
representing samples and columns representing genes.  Let $Z$ be an
$n\times K$ binary matrix with $z_{ik}=1$ if molecular driver $k$
presents in sample $i$, and 0 otherwise. 
That is, the $k$-th column $\bz_k$ defines the subset 
$G_k = \{i:\; z_{ik}=1\}$ of samples that
are linked with the $k$-th molecular driver.
The entire matrix $Z$ defines a multiset $\{G_k,\; k=1,\ldots,K\}$.
Such multisets are known as feature allocation 
\citep{Broderick:13} and are popular tools in machine learning to
implement inference about overlapping subsets of experimental units
(customers etc.).
The special case of non-overlapping subsets that cover all samples,
that is, $G_k \cap G_{\ell}=\emptyset$ and $\bigcup G_k =
\{1,\ldots,n\}$, is a partition. 
See \cite{Broderick:STS:13} for a recent review.
We  use the feature allocation matrix $Z$ to construct a sampling
model for the breast cancer gene expression data $Y$: 
\begin{equation}
   \bY = Z\bbeta + E, 
   \label{eq:FAlikl}
\end{equation}
where $\bbeta$ is a $K\times S$ loading matrix with each entry
$\beta_{kj}$ weighing the contribution of gene $j$ to the driver $k$ and
$E=[e_{ij}]$ is an error matrix with $e_{ij} \sim N(0,\sigma^2)$,
independently. This defines a sampling model for the observed gene
expressions $\bY$ in terms of assumed latent structure $Z$.
That is,
$
  y_{ij} \sim N\left(\sum_{k=1}^Kz_{ik}\beta_{kj}, \sigma^2\right).
$

The key assumption in a feature allocation model
is the prior model on $Z$.
A technically convenient and traditional prior 
is the Indian buffet process (IBP) \citep{griffiths2005infinite}.
One of the key properties of the IBP, in the context of this
application, is the implied independence across columns of the binary
matrix (re-arranging columns in left ordered form or by other
constraints introduces a trivial form of dependence).
This independence is undesirable for the desired inference on
molecular drivers. In particular, independence across columns
 implies a positive prior probability for identical columns, which is
meaningless in the interpretation of columns as distinct molecular
drivers. 

\noindent {\bf{DPP feature allocation.}}
In contrast to the IBP, a DPP prior on the columns $\bz_k$ formalizes
the desired parsimony in identifying latent molecular drivers.
We assume
\begin{equation}
  \{\bz_k,\; k=1,\ldots,K\}  \sim  \DPP(C)
  \mbox{ with }
  C(\bz_\ell, \bz_{\ell'}) = \exp\left\{-
    \frac{\sum_{i=1}^n(z_{i\ell'}-z_{i\ell})^2}{\theta^2}  \right\}.
  \label{eq:FL}
\end{equation}
With  large $n$, 
there is no effective way to
decompose the kernel matrix $C$ ($N\times N$ matrix with $N=2^n$) and to compute the eigenvalues and the
corresponding eigenvectors. We therefore fix $\th$ in
\eqref{eq:FL} and complete the model with 
a conditionally conjugate prior on the coefficients, 
$ \beta_{kj} \iid N(0, \tau^2)$,
and  hyperpriors
$
   1/\sigma^2\sim \Ga(a_0, b_0),
   1/\tau^2\sim \Ga(a_1, b_1).
$

\noindent {\bf{DPP-$K$ feature allocation.}}
In the upcoming 
applications we find it convenient to work with a slight
variation of model \eqref{eq:FL}.
Let $\DPP_K(C)$ denote a DPP prior restricted to a
fixed number of atoms, $K$, and define
\begin{equation}
  \{\bz_k,\; k=1,\ldots,K\} {\mid K} \sim \DPP_K(C)
  \mbox{ and }
  K \sim p(K)
\label{eq:FL-K}
\end{equation}
for some prior $p(K)$.
We refer to the model as the DPP-$K$ feature allocation model.
The reason for introducing the DPP-$K$ model is that it facilitates a
computationally efficient posterior simulation scheme.
Under \eqref{eq:FL} we require a RJ type implementation, following the
general scheme in Section \ref{sec:mcmc}.
However, in some applications it  is difficult to construct proposal distributions that lead to
reasonably mixing Markov chains. Instead we propose in Web Appendix C 
an alternative MCMC scheme under \eqref{eq:FL-K}, which we find to
work well for applications with feature allocation problems.
Posterior inference in model \eqref{eq:FL} as well as in \eqref{eq:FL-K}
is implemented by MCMC posterior simulation.
Details are shown in Web Appendix C.  

\begin{figure}[!h]
\centering
\begin{tabular}{ccc}
  \includegraphics[scale=0.28]{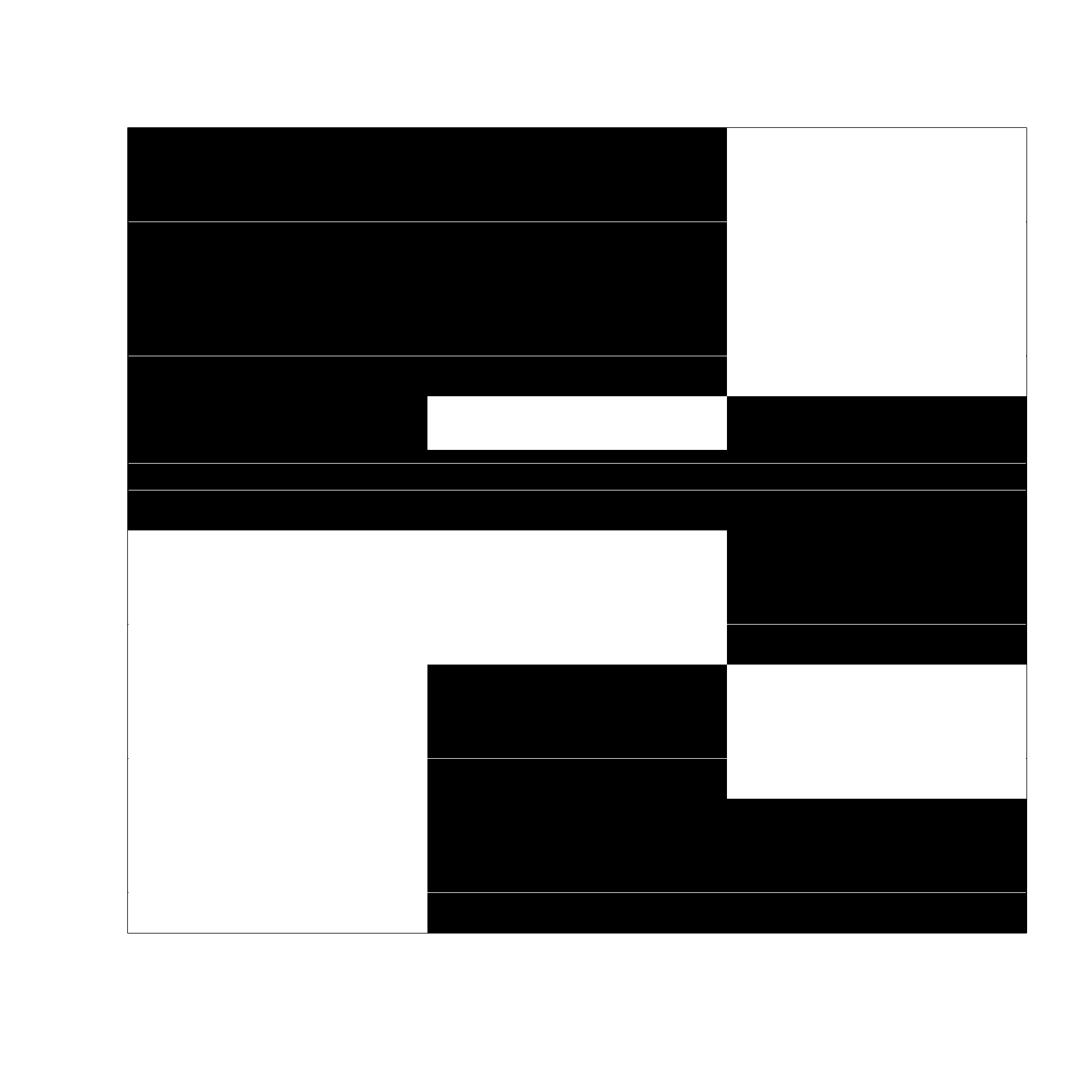}&\includegraphics[scale=0.28]{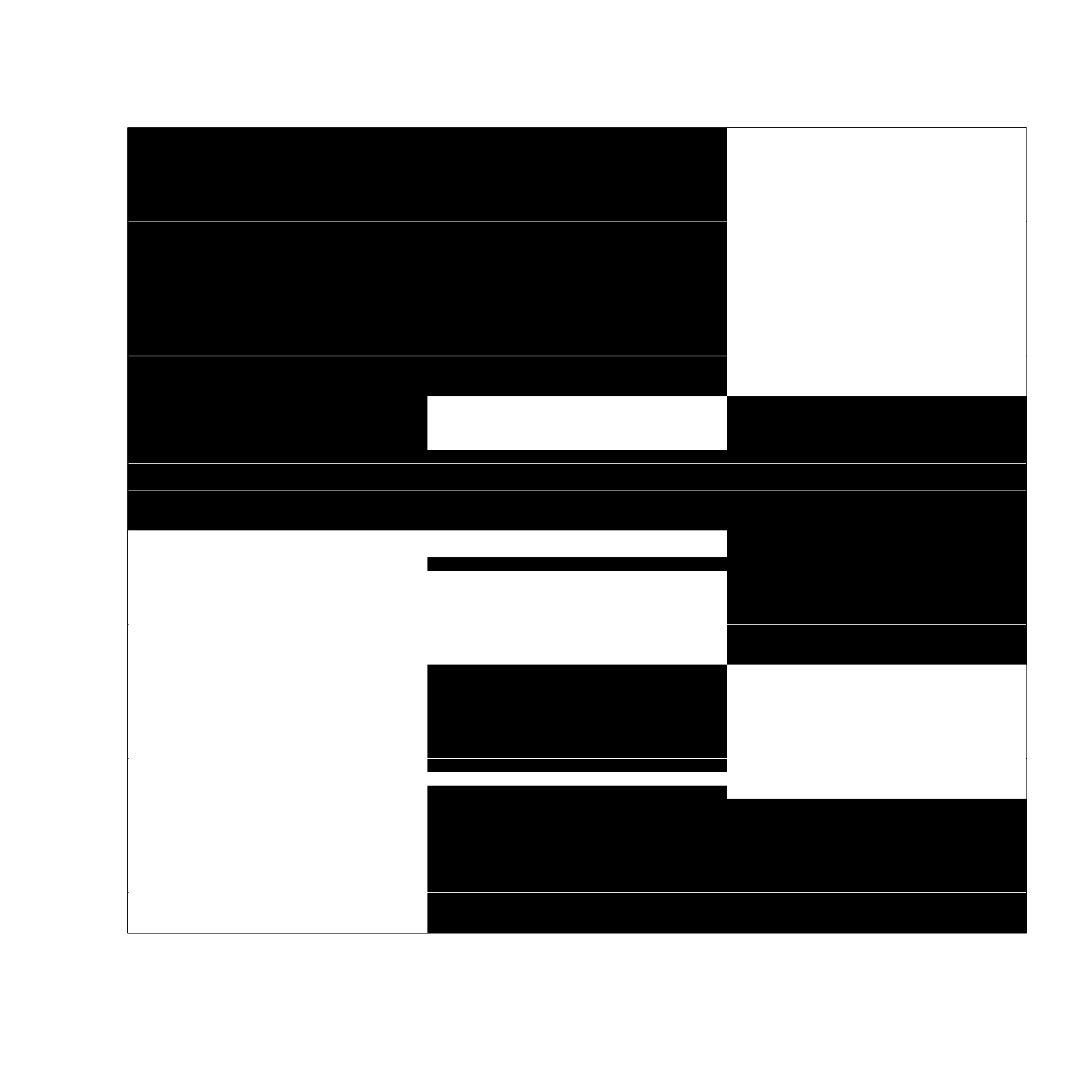} &\includegraphics[scale=0.28]{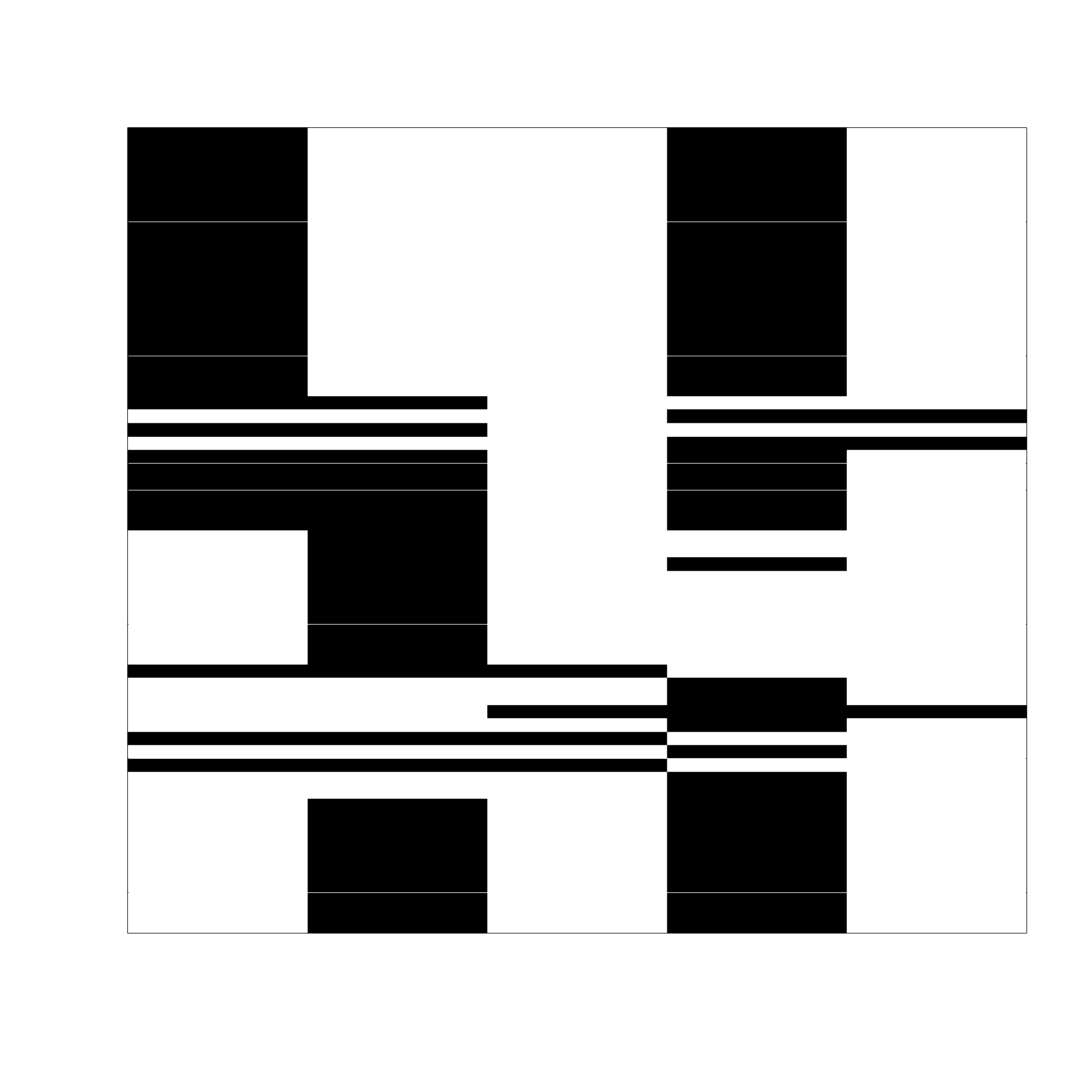} \\
  (a) True $Z^o$ & (b) Estimated $\hat{Z}$ by DPP &(c) Estimated $\hat{Z}$ by IBP\\
  \includegraphics[scale=0.28]{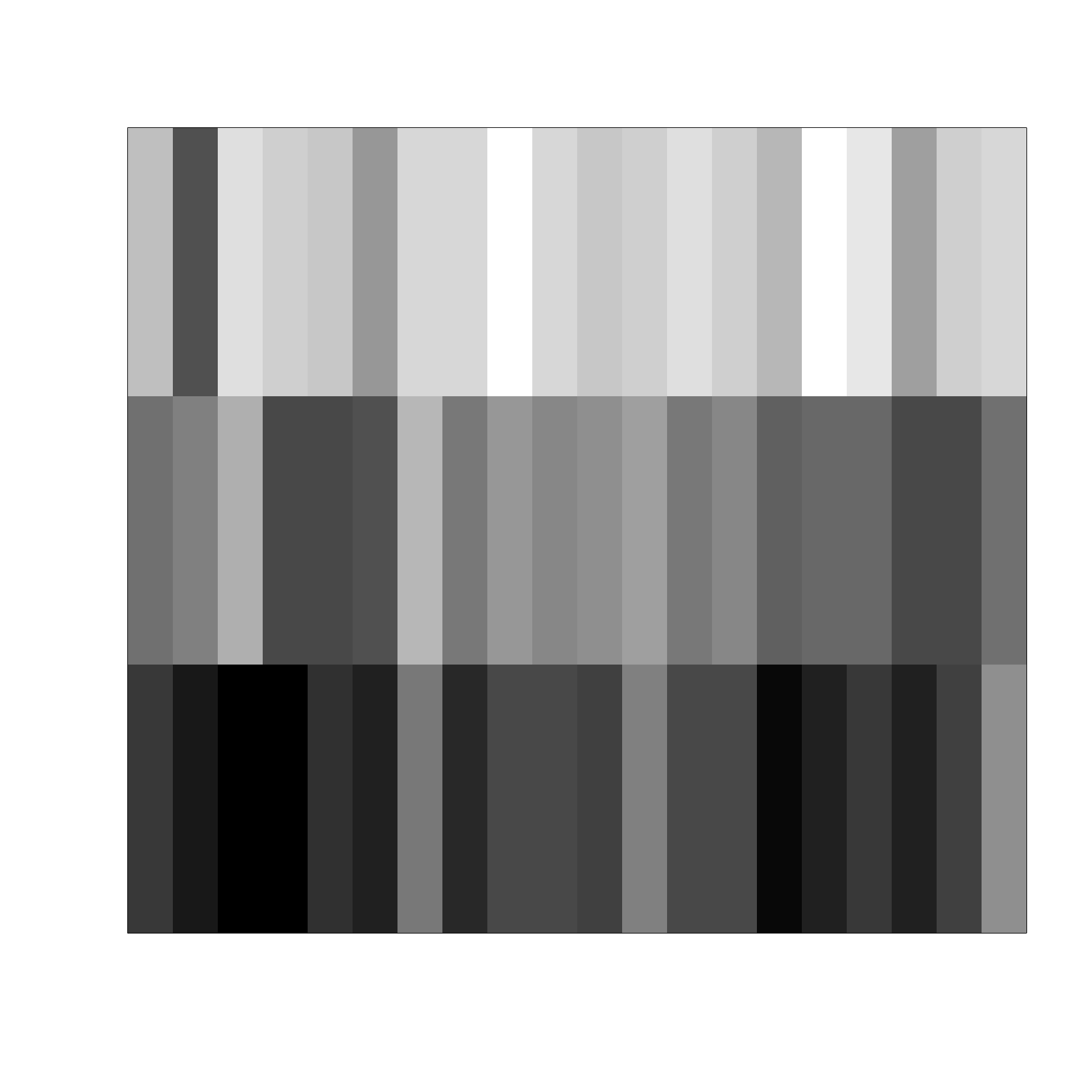}&\includegraphics[scale=0.28]{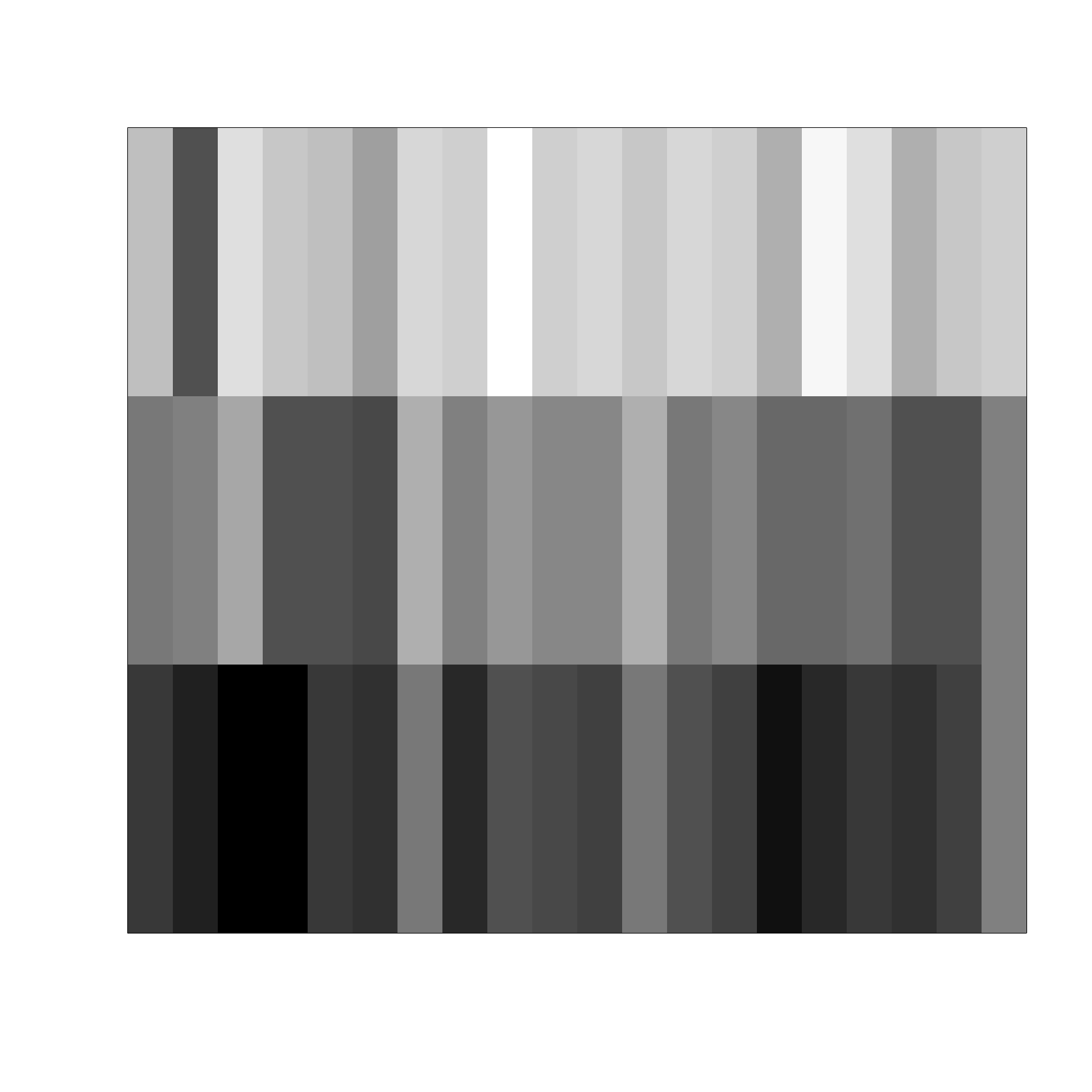} &\includegraphics[scale=0.28]{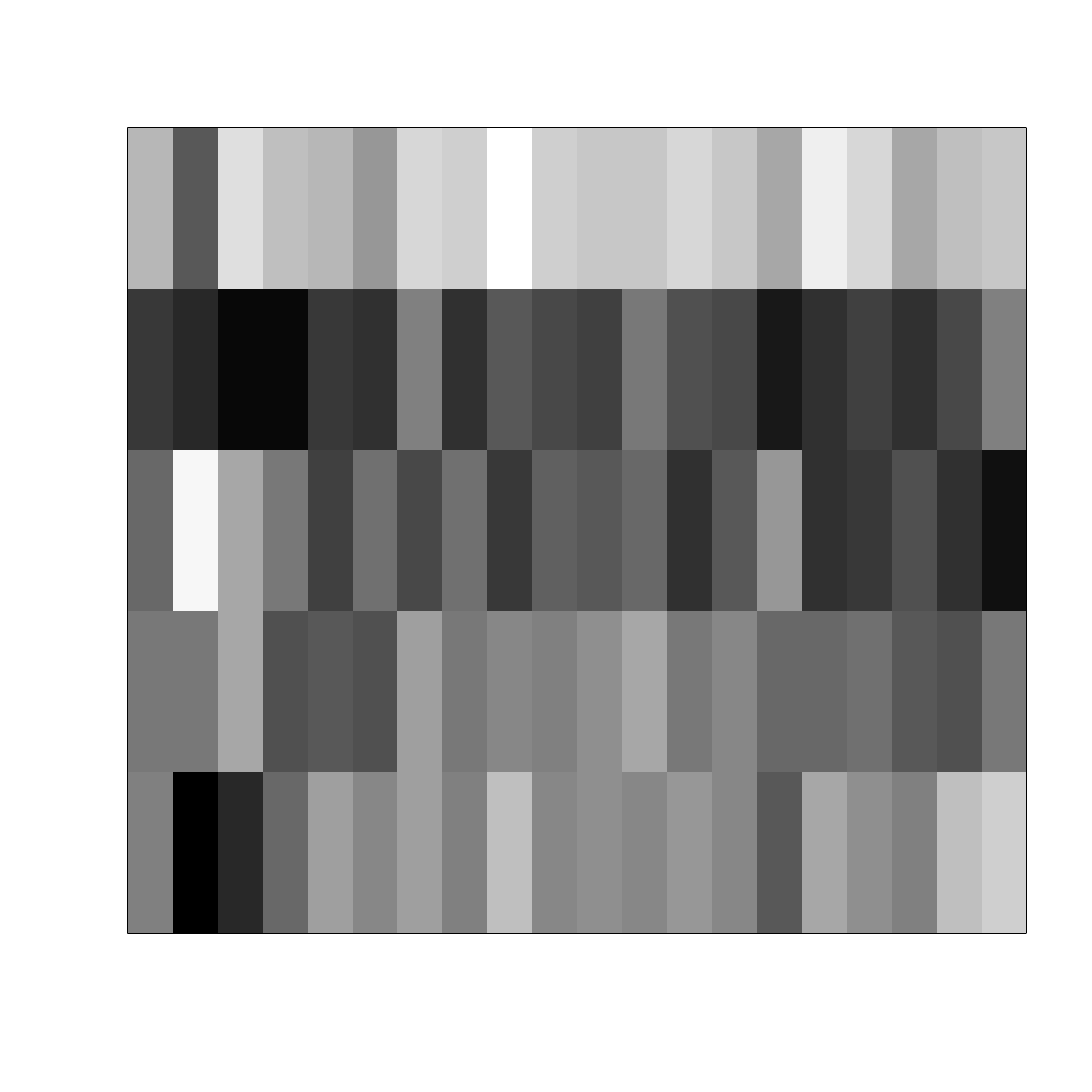} \\
  (d) True $\bbeta^o$ & (e) Estimated $\hat{\bbeta}$ by DPP &(f) Estimated $\hat{\bbeta}$ by IBP\\
  &\includegraphics[scale=0.28]{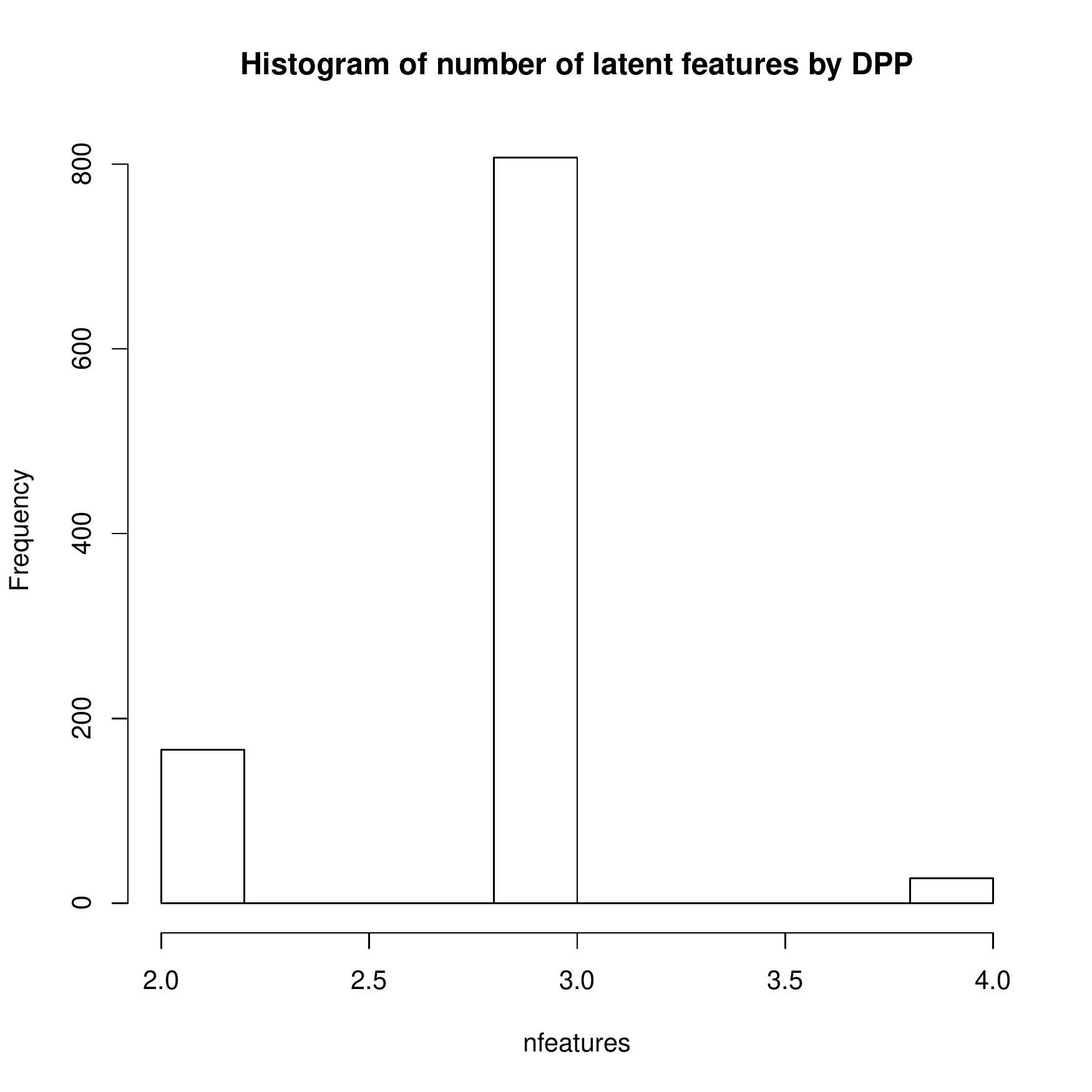} &\includegraphics[scale=0.28]{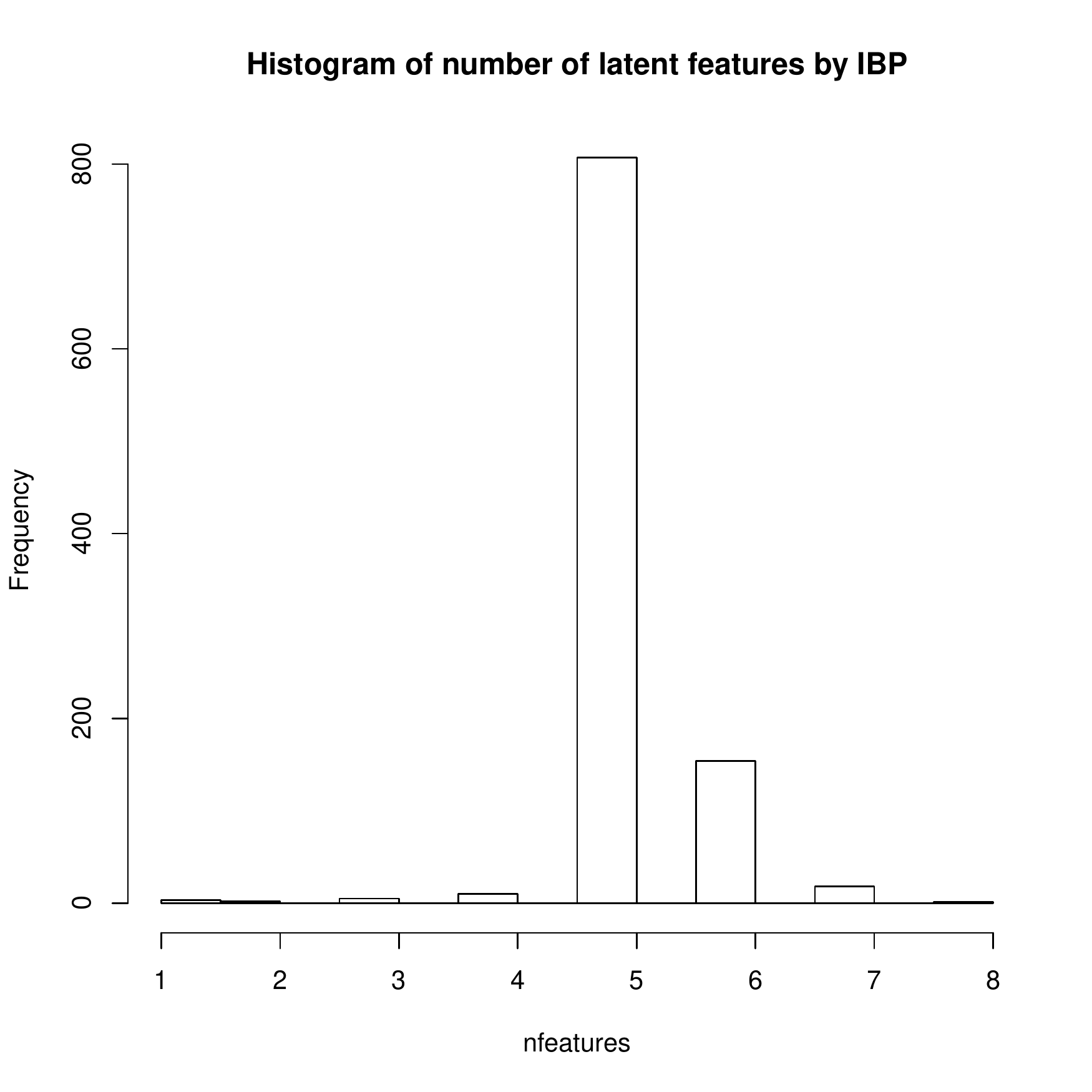} \\
  &(g)&(h)\\
\end{tabular}
\caption{Simulation: DPP feature allocation model.
  Panels (a-c) show the true feature allocation matrix $Z^o$
  and the estimate  $\hat{Z}$ under the DPP prior and the IBP prior,
  respectively. 
  Panels (d-f) show the true feature mean $\bbeta^o$ and the estimated
  $\hat{\bbeta}$ under the DPP prior and under the IBP prior,
  respectively. 
  Panels (g-h) are histograms of the number of 
  latent features identified under DPP prior and IBP prior,
  respectively. } 
\label{fig:simu3}
\end{figure}

\paragraph*{Simulation study.}
We carry out a simulation study to compare inference
under the DPP-$K$ feature allocation prior versus the IBP prior. 
Results are summarized in Figure \ref{fig:simu3}.
More discussion of the results and details of the simulation setup 
are in Web Appendix D. 
With fewer latent features,
inference under the DPP prior can better and more parsimoniously
recover the simulation truth than under a standard IBP prior in this
simulated example.

\subsection{Breast Cancer (BRCA) Data Analysis}

We analyze the TCGA BRCA mRNA expression data
\citep{cancer2012comprehensive}. We focus on $n=150$ tumor samples
classified as basal-like, HER2-enriched (HER2) and luminal A (LumA)
subtypes by PAM50, a well-established 50-gene signature for
distinguishing the gene expression-based ``intrinsic" subtypes of
breast cancer \citep{parker2009supervised}. 
Among those three subtypes, the HER2-enriched subtype is
well studied. 
There are effective therapeutic drugs developed for targeting HER2
breast cancer. 
The basal-like subtype (also known as triple-negative breast cancer
due to its lacking of expression of estrogen receptor (ER),
progesterone receptor (PR) and HER2), and the LumA subtype, which is
known to have lowest overall mutation rate, are poorly understood. As
a result, there is currently no effective targeted therapy for these
two subtypes, leaving chemotherapy as the main therapeutic treatment.
A better characterization of basal-like and LumA subtypes at the
molecular level is 
needed for clinical studies.

\begin{figure}[!h]
\includegraphics[scale=0.6]{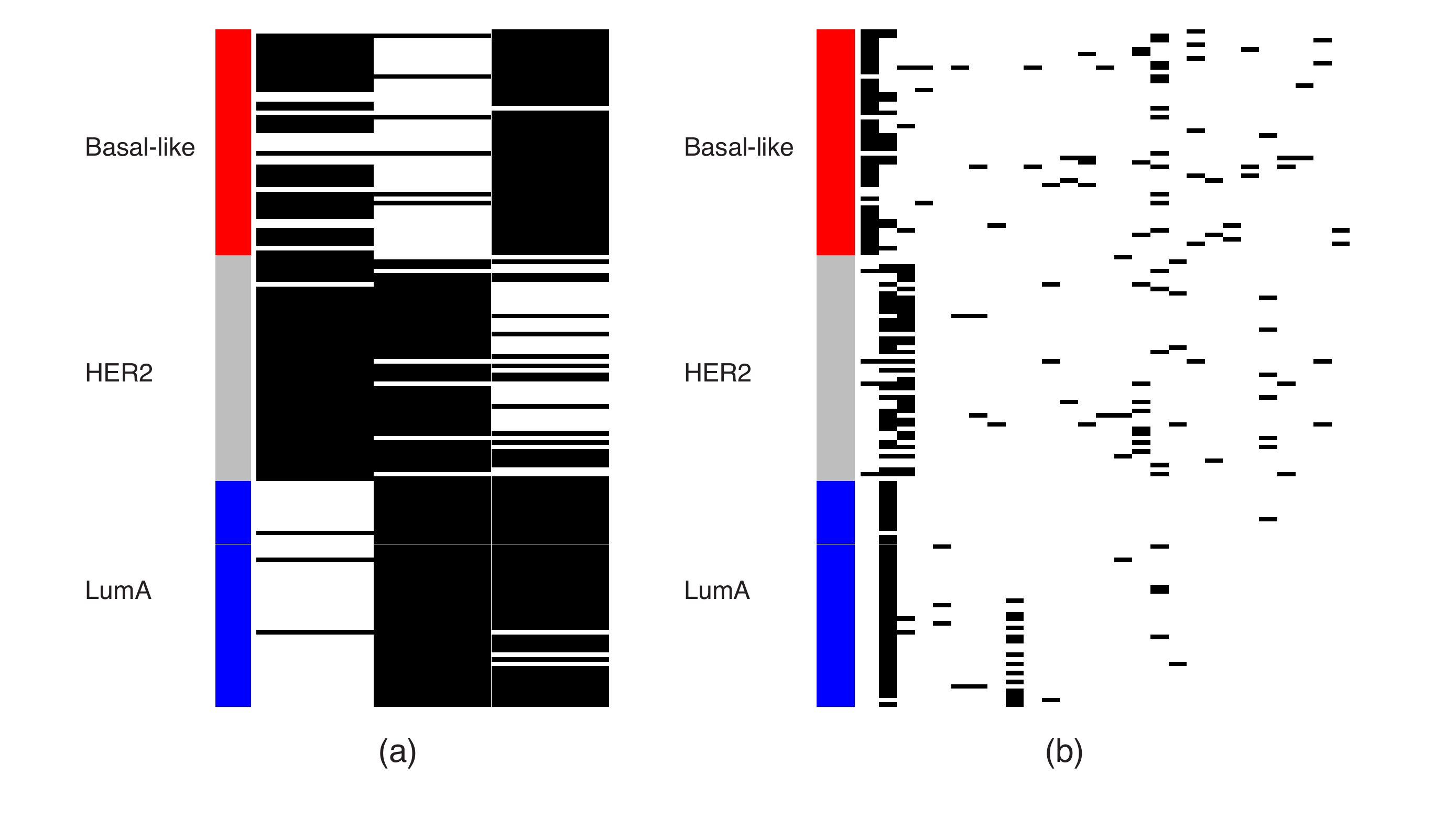}
\caption{BRCA data. 
  Estimated feature allocation matrix $\hat{Z}$
  under the DPP prior (panel a) and under the IBP prior (b).}
\label{fig:real3}
\end{figure}

We implement inference under the proposed
 DPP-$K$ 
 latent feature model and identify $K=3$ latent features.
Figure \ref{fig:real3}a shows the posterior inferred latent feature matrix
$Z$ with different breast cancer subtypes samples marked by different
colors on the left. The basal-like, HER2 and LumA samples show clear
and distinct patterns: 
 35 of 50 basal-like samples exhibit the first and third
features and 44 of 50 are depleted with respect to the second feature; 
43 of 50 HER2 samples exhibit the first two features and 33 of 50 are
depleted with respect to the third feature; 48 of 50 LumA samples
exhibit the second and third features and 47 of 50 are depleted with
the first feature. More biological findings are discussed in Web Appendix E.

For comparison, we analyze the same BRCA dataset under a model 
with an IBP prior. 
It identifies 27 latent features, of which 17 are active in less than
4 samples. Figure \ref{fig:real3}b shows the estimated latent
feature matrix $Z$. The first three features identified under the IBP prior
can distinguish different breast cancer subtypes: 44 of 50 basal-like
samples exhibit the first feature; 4 of 50 HER2 samples and none of
LumA samples exhibit the first feature; 48 of 50 LumA samples exhibit
the third feature. However, for the remaining 24 features, we can not
observe any pattern for different breast cancer subtypes: these
features were sparsely scattered across all samples. 
This is a good example of how the independent prior across features,
as it is implied in the IBP model, leads to a lack of parsimony and
difficult interpretability in the latent structure.
In summary, the DPP prior model provides a less complicated
representation and more interpretable features than
 inference under the  IBP prior model.

%
%

\section{Conclusions}
We argue 
for the use of repulsive priors in models that involve
latent structure as the main inference target. In many such problems
interpretation of the imputed latent structure favors diverse and
parsimonious choices. 
We specifically discuss examples involving inference for mixture models
and 
feature allocation models. In these settings,
commonly used models  assume independence across
latent clusters, features etc., which is technically convenient, but
often inappropriate for the desired inference.
We instead propose the use of DPP models as repulsive priors.
The DPP model is attractive mainly because of the availability of easy
to implement posterior simulation schemes.

We compare inference using DPP priors with standard Bayesian nonparametric
priors in the cluster analysis of renal clear cell carcinoma and a feature allocation analysis
of TCGA BRCA mRNA expression data. Our examples show that using DPP
priors leads to posterior inference that gains substantially in
parsimony and interpretability. Our case study results are methodologically corroborated by our
analysis of  the inferred structures.
Also, DPP priors lead to  a noticeable reduction in model uncertainty and,
consequently, significantly more efficient estimators of latent
structures.

Beyond mixture models and feature allocation models,
inference for latent structures arises naturally in many other
biomedical applications. 
One class of such examples are applications that involve nested
clustering, that is, 
clustering of one set of experimental units (e.g., proteins) with
respect to shared nested partitions on another set of experimental
units (e.g., patients). 
\cite{juhee:13} discussed
such applications, but with independent priors across distinct nested
partitions. 
 In some contexts, 
the latent 
structure of interest could be a graph, for example, a
conditional independence graph that might be shared across some
subpopulations. For example, \cite{riten:15} considered dependence
structure of histone modifications across different conditions.

Several important limitations remain. In some applications repulsive
priors are inappropriate. For example, inference for tumor
heterogeneity might involve a prior across latent hypothetical
subclones. However, following the notion of a phylogenetic tree of
tumor cell subpopulations, some of these latent subclones should
differ by few features (mutations, copy number variations etc.) only. 
Also, important computational limitations may be encountered, depending on 
specific applications. For example, in big data settings, fast posterior approximations developed
for standard prior models may not extend directly to the case of DPP priors \citep{xu2015mad}. 
Finally, problems related to label-switching 
\citep{jasra2005} remain an issue like in any
mixture model. This is the case because the DPP prior remains
exchangeable, for example across the $\mu_k$ in the mixture model
\eqref{eq:prior}.


\section{Supplementary Materials}
Web appendices and figures referenced in Sections 4.1, 5.1 and 5.2 are
available with this paper at the {\it Biometrics} website on Wiley Online
Library. 

\section*{Acknowledgements}
Peter M\"uller and Yanxun Xu's research is partly supported by NIH grant R01 CA132897.

\bibliographystyle{apalike}
\bibliography{DPP}

\begin{thebibliography}{}

\bibitem[Affandi et~al., 2013]{affandi2013approximate}
Affandi, R.~H., Fox, E., and Taskar, B. (2013).
\newblock Approximate inference in continuous determinantal processes.
\newblock In {\em Advances in Neural Information Processing Systems}, pages
  1430--1438.

\bibitem[Broderick et~al., 2013a]{Broderick:STS:13}
Broderick, T., Jordan, M.~I., Pitman, J., et~al. (2013a).
\newblock Cluster and feature modeling from combinatorial stochastic processes.
\newblock {\em Statistical Science}, 28(3):289--312.

\bibitem[Broderick et~al., 2013b]{Broderick:13}
Broderick, T., Pitman, J., Jordan, M.~I., et~al. (2013b).
\newblock Feature allocations, probability functions, and paintboxes.
\newblock {\em Bayesian Analysis}, 8(4):801--836.

\bibitem[Chen et~al., 2013]{chen2013phylogenetic}
Chen, M., Gao, C., and Zhao, H. (2013).
\newblock Phylogenetic {Indian} buffet process: Theory and applications in
  integrative analysis of cancer genomics.
\newblock {\em arXiv preprint arXiv:1307.8229}.

\bibitem[Cocosco et~al., 1997]{cocosco1997brainweb}
Cocosco, C.~A., Kollokian, V., Kwan, R. K.-S., Pike, G.~B., and Evans, A.~C.
  (1997).
\newblock Brainweb: Online interface to a {3D MRI} simulated brain database.
\newblock In {\em NeuroImage}, volume~5, page 425. Citeseer.

\bibitem[DeCarli et~al., 1992]{decarli1992method}
DeCarli, C., Maisog, J., Murphy, D.~G., Teichberg, D., Rapoport, S.~I., and
  Horwitz, B. (1992).
\newblock Method for quantification of brain, ventricular, and subarachnoid
  {CSF} volumes from {MR} images.
\newblock {\em Journal of Computer Assisted Tomography}, 16(2):274--284.

\bibitem[Geyer and M{\o}ller, 1994]{Geyer&Moller94}
Geyer, C.~J. and M{\o}ller, J. (1994).
\newblock {Simulation Procedures and Likelihood Inference for Spatial Point
  Processes}.
\newblock {\em Scandinavian Journal of Statistics}, 21(4):359--373.

\bibitem[Ghahramani and Griffiths, 2006]{griffiths2005infinite}
Ghahramani, Z. and Griffiths, T.~L. (2006).
\newblock Infinite latent feature models and the {I}ndian buffet process.
\newblock In {\em Advances in Neural Information Processing Systems}, pages
  475--482.

\bibitem[Ghoshal, 2010]{Ghoshal:10}
Ghoshal, S. (2010).
\newblock The {D}irichlet process, related priors and posterior asymptotics.
\newblock In Hjort, N.~L., Holmes, C., M\"uller, P., and Walker, S.~G.,
  editors, {\em {B}ayesian Nonparametrics}, pages 22--34. Cambridge University
  Press.

\bibitem[Green, 1995]{green:95}
Green, P.~J. (1995).
\newblock {Reversible jump Markov chain Monte Carlo computation and {B}ayesian
  model determination}.
\newblock {\em Biometrika}, 82:711--732.

\bibitem[Jasra et~al., 2005]{jasra2005}
Jasra, A., Holmes, C.~C., and Stephens, D.~A. (2005).
\newblock Markov chain {M}onte {C}arlo methods and the label switching problem
  in {B}ayesian mixture modeling.
\newblock {\em Statistical Science}, 20(1):50--67.

\bibitem[Kingman, 1992]{kingman:92}
Kingman, J. F.~C. (1992).
\newblock {\em Poisson Processes}.
\newblock Oxford University Press.

\bibitem[Kulesza and Taskar, 2010]{kulesza2010structured}
Kulesza, A. and Taskar, B. (2010).
\newblock Structured determinantal point processes.
\newblock In {\em Advances in Neural Information Processing Systems}, pages
  1171--1179.

\bibitem[Kulesza and Taskar, 2012]{kulesza2012determinantal}
Kulesza, A. and Taskar, B. (2012).
\newblock Determinantal point processes for machine learning.
\newblock {\em Machine Learning}, 5(2-3):123--286.

\bibitem[Kwok and Adams, 2012]{kwok2012priors}
Kwok, J.~T. and Adams, R.~P. (2012).
\newblock Priors for diversity in generative latent variable models.
\newblock In {\em Advances in Neural Information Processing Systems}, pages
  2996--3004.

\bibitem[Lavancier et~al., 2015]{lavancier:15}
Lavancier, F., M{\o}ller, J., and Rubak, E. (2015).
\newblock Determinantal point process models and statistical inference.
\newblock {\em Journal of the Royal Statistical Society: Series B (Statistical
  Methodology)}, 77(4):853--877.

\bibitem[Lee et~al., 2013]{juhee:13}
Lee, J., M\"uller, P., Zhu, Y., and Ji., Y. (2013).
\newblock A nonparametric {B}ayesian model for local clustering.
\newblock {\em Journal of the American Statistics Association}, 108:775--788.

\bibitem[Macchi, 1975]{macchi1975coincidence}
Macchi, O. (1975).
\newblock The coincidence approach to stochastic point processes.
\newblock {\em Advances in Applied Probability}, pages 83--122.

\bibitem[Mitra et~al., 2015]{riten:15}
Mitra, R., M\"uller, P., and Ji, Y. (2015).
\newblock Bayesian graphical models for differential pathways.
\newblock {\em Bayesian Analysis}, to appear.

\bibitem[Parker et~al., 2009]{parker2009supervised}
Parker, J.~S., Mullins, M., Cheang, M.~C., Leung, S., Voduc, D., et~al. (2009).
\newblock Supervised risk predictor of breast cancer based on intrinsic
  subtypes.
\newblock {\em Journal of Clinical Oncology}, 27(8):1160--1167.

\bibitem[Petralia et~al., 2012]{petralia2012repulsive}
Petralia, F., Rao, V., and Dunson, D.~B. (2012).
\newblock Repulsive mixtures.
\newblock In {\em Advances in Neural Information Processing Systems}, pages
  1889--1897.

\bibitem[Rousseau and Mengersen, 2011]{rousseau2011asymptotic}
Rousseau, J. and Mengersen, K. (2011).
\newblock Asymptotic behaviour of the posterior distribution in overfitted
  mixture models.
\newblock {\em Journal of the Royal Statistical Society: Series B (Statistical
  Methodology)}, 73(5):689--710.

\bibitem[{The Cancer Genome Atlas Network} et~al.,
  2012]{cancer2012comprehensive}
{The Cancer Genome Atlas Network} et~al. (2012).
\newblock Comprehensive molecular portraits of human breast tumours.
\newblock {\em Nature}, 490(7418):61--70.

\bibitem[Xu et~al., 2015]{xu2015mad}
Xu, Y., M{\"u}ller, P., Yuan, Y., Gulukota, K., and Ji, Y. (2015).
\newblock {MAD} {B}ayes for tumor heterogeneity -- feature allocation with
  exponential family sampling.
\newblock {\em Journal of the American Statistical Association},
  110(510):503--514.

\bibitem[Yuan et~al., 2014]{yuan2014assessing}
Yuan, Y., Van~Allen, E.~M., Omberg, L., Wagle, N., et~al. (2014).
\newblock Assessing the clinical utility of cancer genomic and proteomic data
  across tumor types.
\newblock {\em Nature Biotechnology}, 32(7):644--652.

\bibitem[Zhu et~al., 1998]{zhu1997gaussian}
Zhu, H., Williams, C., Rohwer, R., and Morciniec, M. (1998).
\newblock Gaussian regression and optimal finite dimensional linear models.
\newblock {\em NATO ASI Series. Series F: Computer and System Sciences}, pages
  167--184.

\end{thebibliography}

\clearpage

\end{document}